\documentclass[oldversion]{aa}  
\usepackage{graphicx}
\usepackage{txfonts}
\usepackage{natbib}
\usepackage{epstopdf}
\bibpunct{(}{)}{;}{a}{}{,} 

\newcommand{\Vir}{\ensuremath{\mu}}
\begin{document}
   \title{A dynamical mass estimator for high z galaxies based on spectroastrometry \thanks{Based on observations collected with European Southern Observatory/Very Large Telescope (ESO/VLT) (proposals 075.A-0300, 076.A-0711 and 178.B-0838), with the Italian TNG, operated by FGG (INAF) at the Spanish Observatorio del Roque de los Muchachos, and with the {\it Spitzer Space Telescope}, operated by JPL (Caltech) under a contract with NASA. }}
   
   \titlerunning{Dynamical mass estimator based on spectroastrometry.}
   \authorrunning{A.Gnerucci}


   \author{A. Gnerucci\inst{1}, A. Marconi\inst{1,2}, G. Cresci\inst{2}, R. Maiolino\inst{3}, F. Mannucci\inst{2}, N.M. F\"orster Schreiber\inst{4}, R.Davies\inst{4}, K. Shapiro\inst{5}, E.K.S. Hicks\inst{6}
          }

   \offprints{gnerucci@arcetri.astro.it}

   \institute{Dipartimento di Fisica e Astronomia, Universit\`a degli Studi di Firenze, Largo E. Fermi 2, 50125 Firenze, Italy\\
              \email{gnerucci@arcetri.astro.it, alessandro.marconi@unifi.it}
              \and
              INAF-Osservatorio Astrofisico di Arcetri, Largo E. Fermi 5, 50125 Firenze, Italy\\
              \email{gcresci,filippo@arcetri.astro.it}
              \and
              INAF-Osservatorio Astronomico di Roma, via di Frascati 33, Monte Porzio Catone, Italy\\
              \email{maiolino@oa-roma.inaf.it}
              \and
              Max-Planck-Institut f\"{u}r extraterrestrische Physik, Giessenbachstrasse, D-85748 Garching, Germany\\
              \email{forster@mpe.mpg.de, davies@mpe.mpg.de}
              \and
              Aerospace Research Laboratories, Northrop Grumman Aerospace Systems, Redondo Beach, CA 90278, USA\\ 
              \email{Kristen.Shapiro@ngc.com}
              \and 
              University of Washington, Department of Astronomy Box 351580, Seattle, WA 98195-1580\\
              \email{ehicks@u.washington.edu}             
             }


 \abstract{
 
   {Galaxy dynamical masses are important physical quantities to constrain galaxy evolutionary models, especially at high redshifts. However, at $z\gtrsim2$ the limited signal to noise ratio and spatial resolution of the data usually do not allow spatially resolved kinematical modeling and very often only virial masses can be estimated from line widths. But even such estimates require a good knowledge of galaxy size, which may be smaller than the spatial resolution. Spectroastrometry is a technique which combines spatial and spectral resolution to probe spatial scales significantly smaller than the spatial resolution of the observations. 
Here we  apply it to the case of high-z galaxies and present a method based on spectroastrometry to estimate dynamical masses of high z galaxies, which overcomes the problem of size determination with poor spatial resolution. 
We construct and calibrate a ``spectroastrometric'' virial mass estimator,  modifying the ``classical'' virial mass formula. 
We apply our method to the [O\,III]  or $\mathrm{H\alpha}$ emission line detected in $z\sim 2-3$ galaxies from AMAZE, LSD and SINS samples and we compare the spectroastrometric estimator with dynamical mass values resulting from full spatially resolved kinematical modeling. The spectroastrometric estimator is found to be a good approximation of dynamical masses, presenting a linear relation with a residual dispersion of only $0.15$~dex. This is a big improvement compared to the ``classical'' virial mass estimator which has a non linear relation and much larger dispersion ( $0.47$~dex ) compared to dynamical masses.
 By applying our calibrated estimator to 16 galaxies from the AMAZE and LSD samples, we obtain masses in the $\sim 10^7-10^{10} M_{\sun}$ range extending the mass range attainable with dynamical modeling.
   }


   }
   \maketitle


\section{Introduction}\label{s1}

The dynamical properties of galaxies have a fundamental role in the context of galaxy formation and evolution. They are a key prediction of theoretical models and are the most direct way to probe the content of dark matter.
In particular, measuring the dynamical mass of a galaxy is the most direct way to constrain the mass of the dark matter haloes, and this quantity is a fundamental prediction of cosmological cold dark matter models of hierarchical structure formation.
In recent years, our observational knowledge in this field has increased enormously, but there remain many dark areas because of the lack of observations that can constrain the complex physical processes involved.
In galaxy evolutionary models (\citealt{Blumenthal:1984}; \citealt{Davis:1985}; \citealt{Springel:2006}, \citealt{Mo:1998}) mergers are believed to play an important role for galaxy formation and evolution but there is also observational evidence for the existence of rotating disks with high star formation rates at $z\sim 2$ (\citealt{Forster-Schreiber:2009}, \citealt{Cresci:2009}).
While there are many dynamical studies on extended samples of $z\sim2$ objects (\citealt{Genzel:2006}; \citealt{Genzel:2008}; \citealt{Forster-Schreiber:2006}; \citealt{Forster-Schreiber:2009};
\citealt{Cresci:2009}; \citealt{Erb:2006}), little is known about the dynamics of galaxies at redshift $z\gtrsim2.5$ and beyond where
only a handful of objects have been observed (\citealt{Nesvadba:2006}; \citealt{Nesvadba:2007}; \citealt{Nesvadba:2008};  \citealt{Jones:2010}; \citealt{Law:2009}; \citealt{Lemoine-Busserolle:2009}; \citealt{Swinbank:2007}; \citealt{Swinbank:2009}). 
The importance of studying  galaxies at very high redshifts ($z\sim 3-4$) and measuring their dynamical masses can be summarised as follows: this redshift range is before the peak of the cosmic star formation rate (see, for example, \citealt{Hopkins:2006}; \citealt{Mannucci:2007}), only a small fraction ($\sim15\%$, \citealt{Pozzetti:2007}) of the stellar mass has been created, it is in this redshift range that the most massive early-type galaxies are expected to form (see, for example, \citealt{Saracco:2003}) and the number of galaxy mergers  is much larger than at later times (\citealt{Conselice:2007}; \citealt{Stewart:2008}). As a consequence the predictions of different models start to diverge significantly at $z\gtrsim2.5$, and this divergence can be very large at $z\sim 3-4$.

We have recently measured dynamical masses in a sizable sample of galaxies at $z\sim 3$ (\citealt{Gnerucci:2011} , \citealt{Cresci:2010}).
One of the lesson learned from our own work and from the literature is that the data for objects at such high redshifts often suffer from poor signal to noise (hereafter S/N) which does not allow spatially resolved kinematical studies and complete dynamical modeling. For this reason one can only estimate the dynamical mass of a galaxy by applying  the virial theorem to its integrated spectrum.

Virial mass estimates are affected by large systematic errors. Apart from those due to the unverified assumption that the system is virialized, one of the principal problems is the estimate of the size of the galaxy, which often suffers from the low intrinsic spatial resolution. In this paper we present an alternative to the classical virial mass estimate based on the technique of spectroastrometry with the aim of obtaining a more accurate dynamical mass estimator. We follow up on the work of \cite{Gnerucci:2010} about the application of the spectroastrometry technique to the study of the dynamics of rotating gas disks.

This paper is based on the data pertaining to two projects focussed on studying metallicity and dynamics of  high-redshift galaxies: AMAZE (Assessing the Mass-Abundance redshift Evolution) (\citealt{Maiolino:2008}, \citealt{Maiolino:2008a}) and LSD (Lyman- break galaxies Stellar populations and Dynamics) (\citealt{Mannucci:2008}, \citealt{Mannucci:2009}). Both projects use integral field spectroscopy of samples of $z\sim3-4$ galaxies in order to derive their chemical and dynamical properties.
In both projects we make use of  data obtained with the Spectrograph for Integral Field Observations in the Near Infrared (SINFONI) at the Very Large Telescope (VLT) of the European Southern Observatory (ESO).
Integral field spectroscopy has proven to be a powerful tool to study galaxy dynamics as it provides two dimensional velocity maps, without the restrictions of longslit studies, which are also plagued by unavoidable light losses and dynamical biases due to a possible misalignment between the slit and the major axis.

The AMAZE sample consists of $\sim30$ galaxies selected with deep Spitzer/IRAC photometry ($3.6-8\mu m$), an important piece of information to derive a reliable stellar mass. These galaxies were observed with SINFONI in seeing-limited mode.
The LSD sample is an unbiased, albeit small, sample of LBGs with available Spitzer and HST imaging.  SINFONI observations were performed with the aid of adaptive optics in order to improve spatial resolution since the aim of this project was to obtain spatially resolved spectra for measuring kinematics and gradients in emission lines.

In \cite{Gnerucci:2011} we performed a kinematical analysis of 33 $z\sim 3$ galaxies from the AMAZE and LSD samples. We found that $\sim30\%$ of the objects show gas kinematics consistent with that of a rotating disk. 
For these ``rotating'' galaxies we performed kinematical modeling by fitting rotating disk models and found dynamical masses in the range $\sim1.8\times10^9M_{\sun}-2.2\times10^{11}M_{\sun}$; we also found that for the majority of the objects the contribution of turbulent motions is important compared to the ordered motions, suggesting that most rotating galaxies at $z\sim3$ are dynamically ÒhotÓ disks. We then built the baryonic Tully-Fisher relation at $z\sim3$. Our measurements indicate that $z\sim3$ galaxies have a stellar-to-dynamical mass ratio smaller than the value in the local universe, confirming the redshift evolution of the relation already found at $z\sim2$. However, the large scatter of the points may also suggest that, at this redshift, the relation is not yet in place, probably due to the young age of the galaxies.

In Sect.~\ref{s2} we briefly summarize observations and data reduction for the data used in this paper. In Sect.~\ref{s3} we describe the principle of the spectroastrometry technique and its application to local black hole mass measurements presented in \cite{Gnerucci:2010}. In Sect.~\ref{s4} we introduce the ``classical'' virial mass estimator and calibrate it by comparing it with the more accurate dynamical mass estimates. In Sect.~\ref{s5} we present our spectroastrometric dynamical mass estimator and in Sect.~\ref{s52} we calibrate it by comparing it with the more accurate dynamic mass measurements. In Sect.~\ref{s6} we apply our method to  galaxies in the AMAZE and LSD samples and discuss the results. Finally, in Sect.~\ref{s7} we draw our conclusions. In Appendix \ref{sa1} we use simulations to understand how our estimator is affected by various dynamical or instrumental parameters.

In the rest of the paper, we will adopt a $\Lambda CDM$ cosmology with $H_0=70 km s^{-1}$, $\Omega_m=0.3$ and $\Omega_\Lambda=0.7$.

\section{The data}\label{s2}

Complete descriptions of the AMAZE and LSD programs, of their sample selection, observations and data reduction are presented in \cite{Maiolino:2008} and \cite{Mannucci:2009}. Here we present a brief summary of the observations and data reduction.

The near-IR spectroscopic observations were obtained using SINFONI, the integral field spectrometer at VLT \citep{Eisenhauer:2003}. For AMAZE galaxies, SINFONI was used in its seeing-limited mode, with the $0.125\arcsec\times0.25\arcsec$ pixel scale and H+K grism, yielding a spectral resolution $R\sim1500$ over the spectral range $1.45-2.41\mu m$. For LSD galaxies, SINFONI was used with the Adaptive Optics module using a bright star close to the galaxy to guide the wavefront correction system.
The (K-band) seeing during the observations was generally better than $0.8\arcsec$.
Data were reduced with the ESO-SINFONI pipeline. The pipeline subtracts the sky from the temporally contiguous frames, flat-fields the images, spectrally calibrates each individual slice and then reconstructs the cube. Individual cubes were aligned in the spatial direction using the offsets of the position of the [O\,III] or $\mathrm{H\alpha}$ line emission peak.
The atmospheric absorption and instrumental response were taken into account and corrected by dividing with a suitable standard star. The gas kinematics used in the following analysis were obtained from [O\,III] $\lambda\lambda\ 4959,5007$\AA\  and $\mathrm{H\alpha}$ for galaxies at $z\sim3$ and $z\sim 2$, respectively.

\subsection{Complete dynamical modeling of the objects}\label{s21}

 In \cite{Gnerucci:2011} we presented the spatially resolved dynamical modeling of the AMAZE and LSD objects. In this paper we will use the results of that modeling as a basis to calibrate the dynamical mass estimator we are developing. We refer to that paper for any detail of the dynamical modeling procedure.

From the AMAZE and LSD samples we selected a subsample of galaxies that show a velocity field consistent with a rotating disk.
In \citealt{Gnerucci:2011} we implemented a simple but quantitative method to identify smooth velocity gradients, which are the signature of a gas kinematics consistent with disk rotation. This method is based on classifying a galaxy as ``rotating'' or ``non rotating'' according to whether its velocity map is well fitted (or not) by a plane. Using this method we selected from the AMAZE and LSD samples 11 ``rotating'' objects for the complete dynamical modeling, but we also identified 6 objects as ``unclassified'' because of the low spatial resolution or S/N (see section 3 and table 2 of \citealt{Gnerucci:2011}).  For the ``rotating'' objects we performed a complete dynamical modeling.

The adopted dynamical model (see \citealt{Gnerucci:2011} for details) assumes that the ionized gas is circularly rotating in a thin disk neglecting all hydrodynamical effects, thus the disk motion is entirely determined by the gravitational potential. The galaxy gravitational potential is generated by the galaxy mass distribution that is modeled by an exponential disk mass distribution.
The velocity along the line of sight for a given position on the sky is then derived by taking into account geometrical projection effects.
The model takes also into account instrumental beam smearing and binning over detector spatial and velocity axes. 

The spatial resolution of the final data cubes was estimated following kinematical considerations and was typically $\sim0.6\arcsec$ (see \citealt{Gnerucci:2011} for details).

The values of the model parameters (e.g. position angle of the disk line of nodes, inclination of the disk, characteristic radius of the disk, dynamical mass) are obtained by fitting this model to the object velocity map. In particular it yields the best fit value for the dynamical mass of the galaxy $M_{dyn}$, defined as the total mass enclosed in a $10$ kpc radius (e.g.~\citealt{Cresci:2009}) and this is the dynamical mass value we will compare to our estimator prediction.
 In table \ref{tab1} we report the dynamical masses of the best fit models for the various objects analyzed in \cite{Gnerucci:2011}.

\section{Virial mass estimator}\label{s4}

  \begin{figure*}[!ht]
  \centering
  \includegraphics[width=0.48\linewidth]{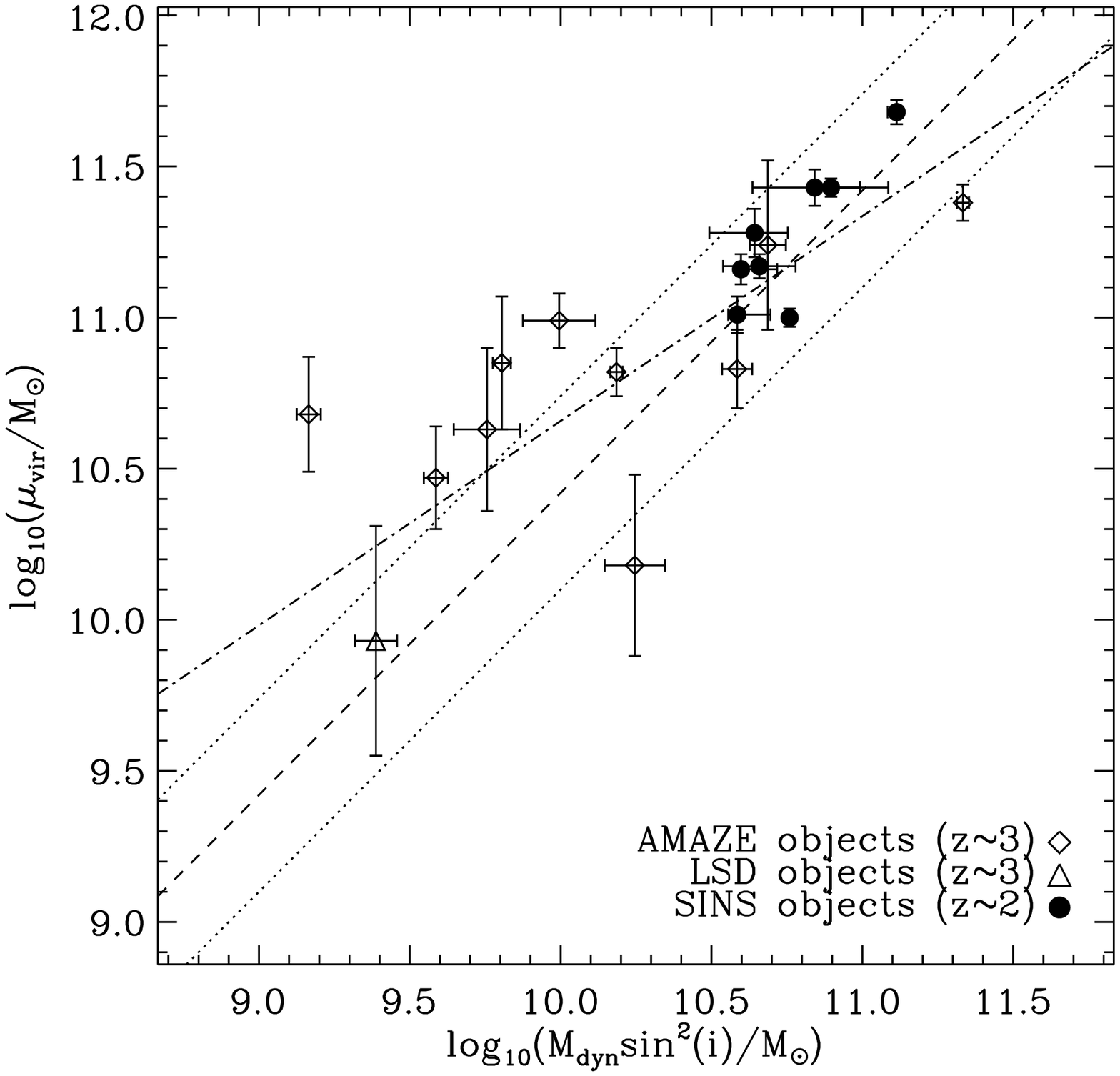}
  \includegraphics[width=0.48\linewidth]{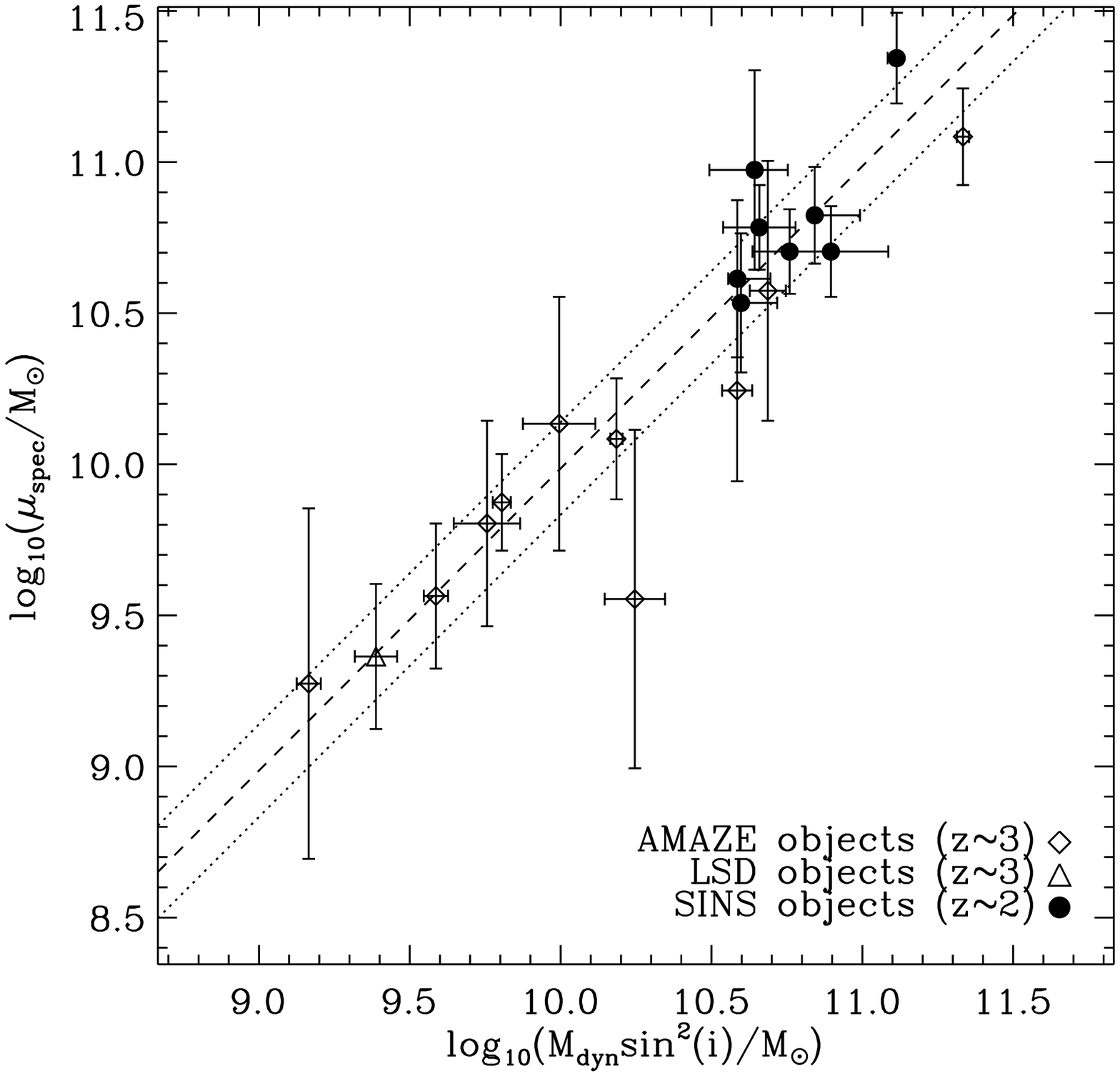}
 \caption{Calibration of the dynamical mass estimator for the AMAZE, LSD and SINS objects. Left panel: calibration of the ``classical'' virial product $\Vir_{vir}=FWHM^2R_c/G$. Dashed line: best-fit linear relation. Dotted lines: rms dispersion of the residuals from the fitted linear relation ($0.3$~dex). Dot-dashed line: best-fit non linear relation. See Sect.~\ref{s4} for details. Right panel: calibration of the ``Spectroastrometric'' virial product $\Vir_{spec}=FWHM^2r_{spec}/G$. Dashed line: best-fit linear relation. Dotted lines: rms dispersion of the residuals from the fitted relation ($0.15$dex). See Sect.~\ref{s52} for details.}
  \label{fig04}
  \end{figure*}

The main and more reliable technique to estimate the dynamical mass of a high redshift galaxy is the modeling of spatially resolved gas kinematics. This technique requires that an object is well spatially resolved and has a high S/N to constrain all the model parameters with sufficient accuracy. Unfortunately these two requirements are not often met for high redshift galaxies, which are in many cases only marginally spatially resolved and with poor S/N. In these cases a method to estimate the dynamical mass of the object is to use the integrated spectrum of the object and  apply the virial theorem.

Applying the virial theorem, the dynamical mass enclosed in a sphere of radius $R_c$ is given by the simple following equation.
\begin{equation}
M(R_c)=f\frac{\Delta V^2\, R_c}{G}
\label{e1}
\end{equation}
where $\Delta V$ is the velocity dispersion along the line of sight which is usually estimated from, e.g., FWHM or velocity dispersion of the line profile or from the observed velocity gradient. The $f$-factor takes into account the geometry of the mass distribution and velocity projection along the line of sight; its actual value also depends on the choice of the velocity estimator.

The choice of the appropriate $f$ is based on estimating its average value over several galaxy mass distribution models, but also depends on how the gas velocity and the galaxy characteristic radius are estimated. Following \cite{Binney:2008}, \cite{Epinat:2009} use a value of $2.25$, an average value of known galactic mass distribution models. Taking into account an average inclination correction, \cite{Erb:2006} obtain a value of $3.4$. Finally, in the case of dispersion-dominated objects, \cite{Forster-Schreiber:2009} adopt an isotropic virial estimate with a value of $6.7$, appropriate for a variety of galaxy mass distributions\footnote{The $f$ values presented here correspond to using the velocity dispersion of the line as an estimate of velocity in eq.~\ref{e1}.}.

The principal problem of the virial mass estimate lies in the estimate of the characteristic dimension $R_c$. First one needs a continuum image of the source with large enough S/N to estimate, e.g., the half light radius. This size has then to be corrected to take into account the finite spatial resolution of the observations; indeed one is often dealing with objects which are marginally spatially resolved and the estimate of $R_c$ can lead to large uncertainties and systematic errors. Second, one needs to estimate the velocity dispersion of the gas from the integrated spectrum. By combining gas velocity dispersion with continuum sizes one has to assume that gas emission is co-spatial with continuum emission and, in general, this assumption cannot be straightforwardly verified. Indeed, there is some disagreement in the few high-z studies regarding the comparison of continuum and emission line morphologies. \cite{Kriek:2009} compared HST/NIC2 and SINFONI $\mathrm{H\alpha}$ morphologies of a spectroscopic sample of massive galaxies at $z\sim2.3$, and  found in many cases significantly different morphologies.
On the other hand \cite{Forster-Schreiber:2011} observed six galaxies at $z\sim2$ with HST/NIC2, finding a good agreement with SINFONI $\mathrm{H\alpha}$ morphologies.

\cite{Forster-Schreiber:2006,Forster-Schreiber:2009} introduce a variation to the classical virial mass estimator by
using the projected circular velocity at the turnover in eq.~\ref{e1}. They estimate this quantity by measuring the amplitude of the observed velocity gradient or the FWHM of the line in the integrated source spectrum. Then they multiply these quantities for an appropriate conversion factor obtained by measuring their average ratio with the projected circular velocity at the turnover in a subsample of galaxies modeled with rotating disks (see also \citealt{Tacconi:2006,Tacconi:2008}).

Using the data from the AMAZE and LSD projects (see Sect.~\ref{s2} for more detailed description) we can calculate the ``classical'' virial mass for several galaxies  (using eq.~\ref{e1} ) and compare it with the more accurate estimates from dynamical modeling of spatially resolved gas kinematics in \cite{Gnerucci:2011} (see Sect.~\ref{s21}). 

In particular we estimate the characteristic radius $R_c$ from the Gaussian HWHM of the emission line image of the galaxy (not having detected any continuum for most objects) correcting this value for the instrumental beam smearing \citep{Bouche:2007}.

In the left panel of Fig.~\ref{fig04} we compare virial products ($\Vir_{vir}=FWHM^2R_c/G$) with masses resulting from full dynamical modeling ($M_{dyn}$) in \cite{Gnerucci:2011}. The numbers plotted in the figure are also reported in table \ref{tab1}. We also extend this comparison to a subsample of $8$ rotating disks from the SINS survey at $z\sim2$ (\citealt {Forster-Schreiber:2006a}, \citealt{Forster-Schreiber:2009}, \citealt{Genzel:2008}, \citealt{Cresci:2009}) selected for their high S/N, ``rotating disk'' gas kinematics and robust dynamical modeling.

For the comparison we do not use the dynamical mass $M_{dyn}$ itself, but ($M_{dyn}sin^2i$), where $i$ is the inclination of the galaxy disk. Due to the coupling of dynamical mass and inclination,  $M_{dyn}sin^2i$ is the quantity which is less affected by systematic errors due to the uncertain disk inclination which is difficult to constrain especially with low  S/N and barely resolved sources (see \cite{Gnerucci:2011}). Moreover, the uncertain disk inclination also affects the virial mass estimate
because the FWHM is determined by the line of sight velocity dispersion.

We note that the virial product $\Vir_{vir}$ and $M_{dyn}sin^2i$ have a non-linear relation with large dispersion which is not the best circumstance for a reliable prediction of $M_{dyn}sin^2i$ from $\Vir_{vir}$. This is to be expected, to some extent, since $\Vir_{vir}$ depends on the line FWHM and $M_{dyn}$ depends on the velocity gradient. Therefore source size and seeing are both going to affect $\Vir_{vir}$ and $M_{dyn}$ in opposite ways : for a given velocity field, a reduction in spatial resolution will increase the spatially unresolved rotation and the line FWHM will get larger while, at the same time, the velocity gradient will be reduced. This might explain the measured slope in Fig.~\ref{fig04}a.

By imposing the expected linear relation  $\log(\Vir_{vir})=\log(M_{dyn}sin^2i)-\log f'$,  the mean rms dispersion of the fit residuals is $0.3$~dex. By letting the slope vary freely, one obtains $\log(\Vir_{vir})=(0.66\pm0.05)\log(M_{dyn}sin^2i)-\log f'$  and the dispersion reduces to $0.21$~dex. The ``classical'' virial mass estimate can be biased by systematic errors principally originating from dimension estimates and therefore it is not a robust proxy for the dynamical mass.

\section{The Spectroastrometric technique}\label{s3}

Spectroastrometry, originally introduced by \cite{beckers:1982}, \cite{christy:1983} and \cite{Aime:1988} to detect unresolved binaries, has been used by several authors to study pre main sequence binaries and the gas disks surrounding pre main sequence stars \citep{Baines:2004, Porter:2004, Porter:2005, takami:2003, Whelan:2005}.
More recently, \cite{Gnerucci:2010} studied the application of spectroastrometry to constrain the kinematics of rotating gas disks in local galactic nuclei and to measure the mass of the putative supermassive black holes using either longslit spectra or integral field spectra.

The fundamental advantage of the spectroastrometric method is that, in principle, it can provide position measurements on scales smaller than the spatial resolution of the observations. This is due to the ability to separate the various spectral features of the source and observe the spatial light distribution of these features separately (see \cite{Gnerucci:2010} for a detailed analysis and discussion).

\subsection{Measuring masses with spectroastrometry}\label{s31}

Here we explain the basics of spectroastrometry and its application to measuring the masses of local Black Holes (hereafter BH), i.e. the application to dynamical studies of  gas disks rotating around point-like mass distributions, as described in \cite{Gnerucci:2010}.

The spectroastrometrical method consists of measuring the photocenter of light emission in different wavelength or velocity channels.
On a longslit spectrum of a continuum subtracted emission line (the so-called position-velocity diagram) the rotation curve denotes the mean gas velocity as a function of the position along the slit whereas the spectroastrometric curve provides the mean position of the emitting gas as a function of velocity (see Fig.~1 of \cite{Gnerucci:2010}). The two curves clearly analyze the same spectrum from complementary points of view.

\cite{Gnerucci:2010} performed an extensive set of simulations in order to understand how the spectroastrometric curve is affected by the object's own dynamical features or by the instrumental setup.
Such simulations show that the presence of a supermassive black hole is revealed by a turn-over in the spectroastrometric curve, with the high velocity component approaching a null spatial offset from the location of the galaxy nucleus. All the relevant information about the BH is encoded in the Òhigh velocityÓ range of the spectroastrometric curve, which is almost independent of the spatial resolution of the observations and of the intrinsic line flux spatial distribution. According to the simulations, the use of spectroastrometry can allow the detection of BHs whose apparent size of the sphere of influence is as small as $\sim1/10$ of the spatial resolution.

\cite{Gnerucci:2010} then provided a simple method to estimate BH masses from spectroastrometric curves. This method consists of the determination of the spectroastrometric map, that is the positions, in each velocity bin, of the emission line photocenters on the plane of the sky. These are obtained by combining longslit spectra centered on a galaxy nucleus at different position angles, but can be trivially obtained from integral field spectra. From the Òhigh velocityÓ points in the spectroastrometric map one can then obtain a rotation curve to easily estimate the BH mass. This method has been tested with simulated data. The test demonstrates the possibility to reconstruct the rotation curve down to radii of $\sim1/10$ of the spatial resolution of the data and, with seeing limited observations ($\sim0. 5\arcsec$), to being able to detect a BH with mass $10^{6.5} M_{\sun}(D/3.5 Mpc)$, where D is the galaxy distance, a factor $\sim10$ better than can be done with the classical method based on rotation curves.

 The $\sim1/10$ gain with respect to the spatial resolution clearly depends on the S/N of the data which ultimately determines the accuracy with which we can estimate photocenters.

In this paper we make use of the results of \cite{Gnerucci:2010} to apply the spectroastrometric technique to the study of the gas dynamics in high redshift galaxies. In particular we use spectroastrometry to improve the ``classical'' virial mass estimates. As noted in the introduction we work on IFU data (SINFONI integral field spectra).

The principal difference with respect to its application to local BHs is that we deal with rotating gas disk dynamics driven by extended, instead of  point-like, mass distributions. But in both cases spatial resolution is an issue, since  the rotating disks are often barely resolved.

\section{Spectroastrometric mass estimator}\label{s5}

The spectroastrometric dynamical mass estimator presented in this paper is based on measuring the characteristic dimension of the object by means of the spectroastrometric technique. As for the classical virial mass estimates we assume that the object is a rotating disk. Following this assumption, we expect that the redshifted gas is located principally on one side of the disk, whereas the blueshifted gas is located on the other side. Therefore, if we ideally obtain images of the object in the ``red'' and  ``blue sides'', these two images have to be spatially shifted due to the rotation. We estimate the characteristic radius of the object by measuring this shift between the red and blue sides with spectroastrometry. In the rest of the paper we will refer to ``sizes'' based on spectroastrometry. This is indeed a new measure of size slightly different from all other measures typically used (e.g.~half-light radius, scale length, HWHM, etc.). This ``size'' is in fact the average distance between the redshifted and blueshifted gas weighted on the surface brightness of line emission. By means of simulations (see Appendix \ref{sa1}) we can assess that for an exponential disk model this ``spectroastrometric size'' correspond on average to a fraction of $\sim0.75$ of the disk scale length, although this relation depends on the assumed disk profile.
 
\begin{figure}[!ht]
  \centering
  \includegraphics[width=0.99\linewidth]{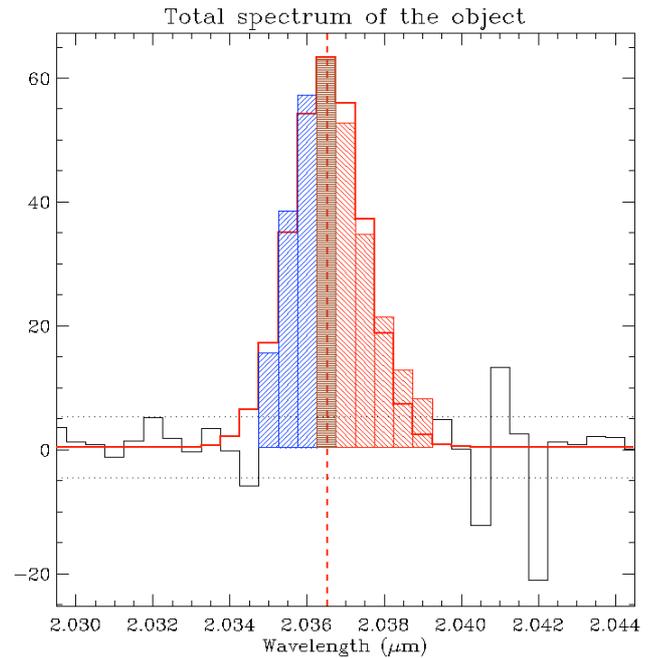}
  \caption{Example of the integrated spectrum for an object ({\it SSA22a-C16}). Overplotted is the fitted model of the line (red continuous line), the central wavelength (red dashed vertical line), the rms of residuals (two dotted horizontal lines) and the selected bins for the blue and red sides (respectively blue and red filled bins) and the central bin (brown filled bin).}
  \label{fig01}
  \end{figure}

The first step is to obtain a spatially integrated spectrum of the source. We then fit the observed line profile with a simple Gaussian function and estimate the line width ($\sigma$ or FWHM).
Using the fitted line profile model we select the ``red'' and the ``blue'' sides of the emission line. The red side is identified by the wavelength range bordered by the central wavelength of the line and the most extreme ``red'' wavelength for which all the following conditions are satisfied: (i) the line model is  greater or equal to $5\%$ of the maximum amplitude, (ii) the data spectrum is greater than or equal to the rms of the fit residuals and (iii) the wavelength is less than $3\sigma$ from the central wavelength. We similarly define the wavelength range for the ``blue'' side of the line (see Fig.~\ref{fig01}). We then collapse all the velocity planes in the red and blue sides of the line to obtain the ``red'' and ``blue'' images of the object, respectively. The spatial plane related to the central wavelength bin is added to the red and blue images but with a weight given by the fraction of the bin lying on the red and blue sides (see Fig.~\ref{fig01}).

  \begin{figure*}[!ht]
  \centering
  \includegraphics[width=0.48\linewidth]{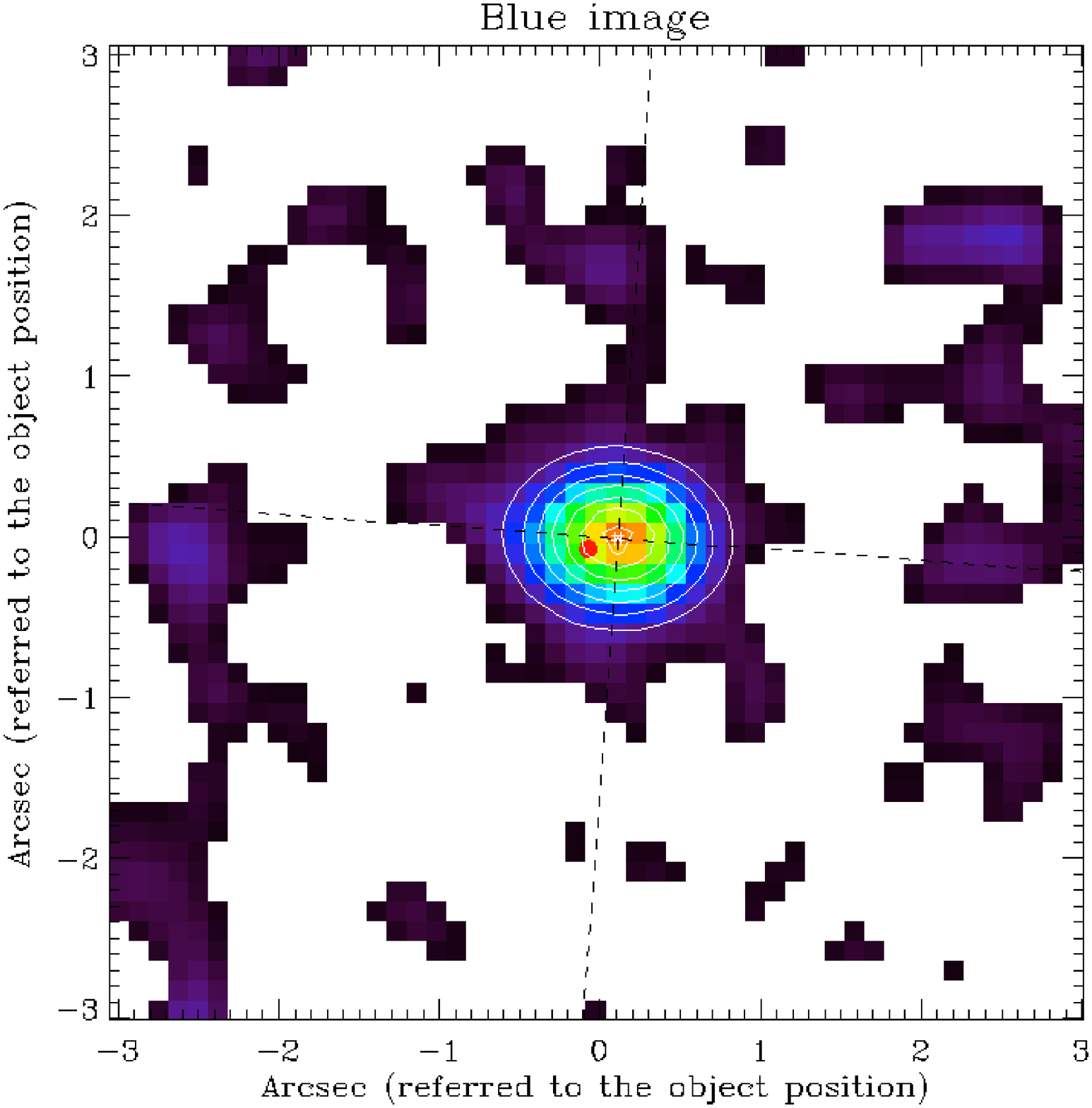}
  \includegraphics[width=0.48\linewidth]{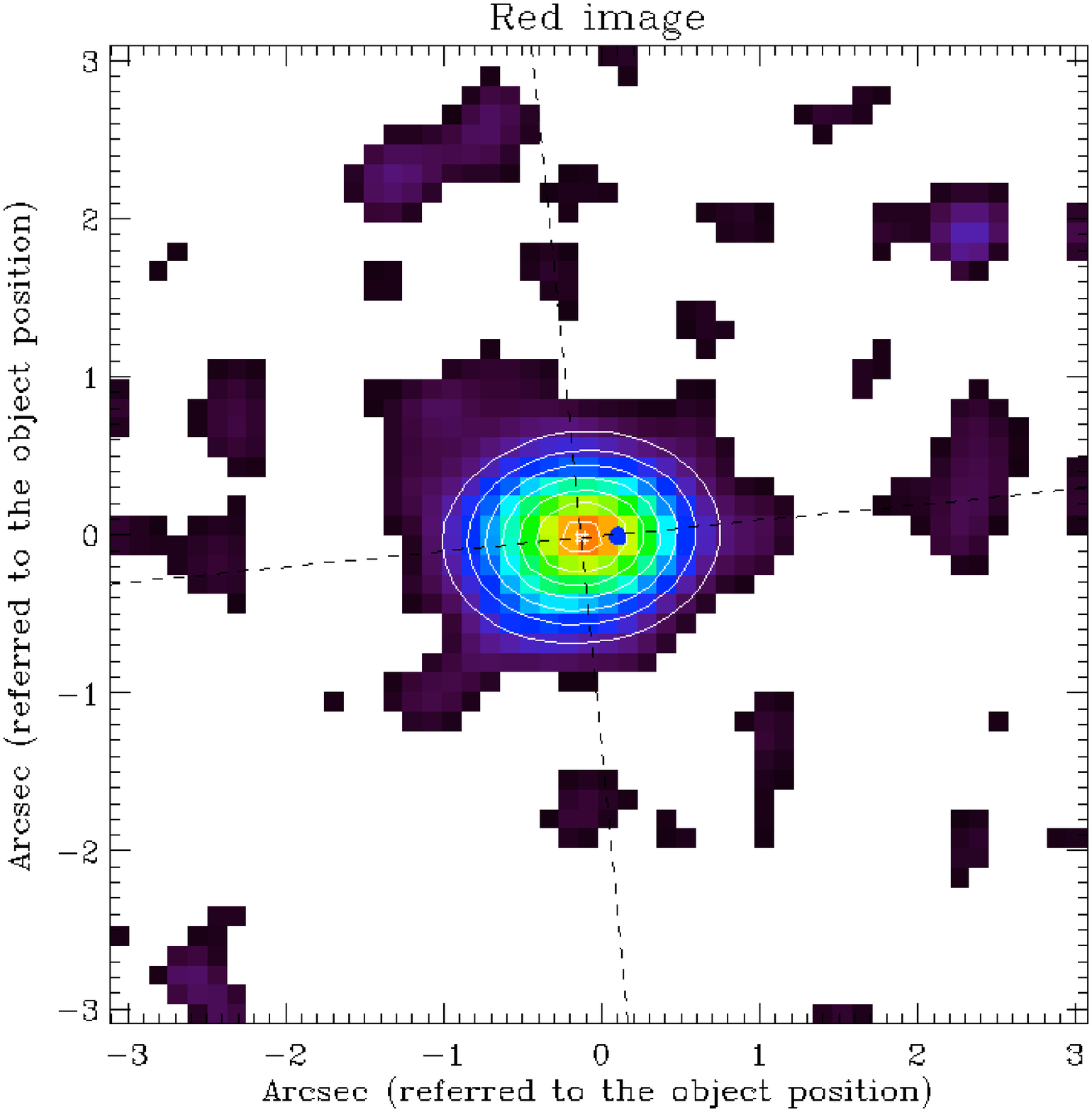}
  \caption{Example of the blue and red image for the same object of Fig.~\ref{fig01}. White isophotes and black dashed lines represent the best model and axis orientation for the two dimensional gaussian function fitted to the images. The blue and red filled circles represent the position of the centroid of respectively the other image.}
  \label{fig02}
  \end{figure*}
  
  \begin{figure}[!ht]
  \centering
  \includegraphics[width=0.99\linewidth]{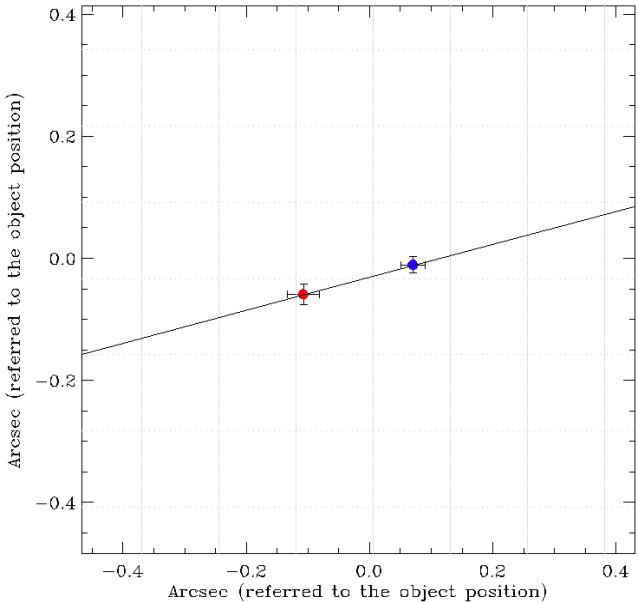}
  \caption{Map of the blue and red centroids for the same example of Figs.~\ref{fig01} and \ref{fig02} The dotted gray grid represents the pixel layout.}
  \label{fig03}
  \end{figure}

The next step is to fit such images with a two-dimensional Gaussian function and determine the position  of the ``blue'' and ``red'' light centroids (see Fig.~\ref{fig03}). We can now define the spectroastrometric characteristic radius ($r_{spec}$) as the half of the distance between the two centroids.
Note that the orientation of the direction connecting the two centroids, in the assumption of a rotating disk kinematics, identifies  the disk line of nodes. Figures \ref{fig01} to \ref{fig03} show as an example the application of the method to the object {\it SSA22a-C16} from the AMAZE sample.

We can understand the physical meaning of the spectroastrometric characteristic radius ($r_{spec}$) in the assumption of  rotating disk kinematics. We obtain the blue and red images by binning the spectral dimension of the data cube, and then measuring the light centroid position for each of the two resulting bins. Therefore we can consider the blue and red centroids, and their associated average velocities, as a ``two-point'' rotation curve of the gas: each point is defined by its velocity ($+\sigma$ and $-\sigma$) and its distance from the center ($+r_{spec}$ and $-r_{spec}$), i.e. from the global peak of line emission.
We note that we are actually measuring the rotation curve of the object using spectroastrometry as in \cite{Gnerucci:2010}, but with the difference that here we are considering only two spectral bins. This is a natural consequence of the poor S/N data we have to deal with.

Finally we can define the spectroastrometric mass estimator ($M_{spec}$) as:
\begin{equation}
M_{spec}sin^2i=f_{spec}\,\Vir_{spec}=f_{spec}\,\frac{FWHM^2\, r_{spec}}{G}
\label{e3}
\end{equation}

where $f_{spec}$ is a calibration factor that takes into account both the geometry of the mass distribution and our choices for measuring velocities and dimensions. The value of  $f_{spec}$ can be calibrated empirically by comparing the virial product with the more accurate masses from full dynamical modeling. We have inserted the factor $sin^2i$ in the first term because of the coupling to the disk inclination discussed above.

The principal advantage of our estimator with respect to the classical virial mass is the way we estimate the characteristic dimension of the galaxy. Firstly, because we estimate it directly using the ionized gas emission, it is more directly related to the observed kinematics than the continuum. In addition, its measurement is not affected by the finite spatial resolution, and we can accurately estimate sizes down to $\sim 1/10$ of the spatial resolution.

As previously observed, this is the principal advantage of the spectroastrometric method and becomes a fundamental feature when dealing with objects observed with poor intrinsic spatial resolution as in the case of high z galaxies. In the example presented in Figs.~\ref{fig01}, \ref{fig02}, and \ref{fig03}, we can estimate the blue and red centroid positions, and consequently $r_{spec}$, with an accuracy of $\sim0.03\arcsec$ that is $\sim1/4$ of the pixel size ($0.125\arcsec$) and $\sim1/20$ of the spatial resolution of the data (for the data used in the example and for most of the data used in this paper the spatial resolution is $\sim0.6\arcsec$ FWHM). With any other method, the estimate of the galaxy half light radius needs to be corrected for beam smearing with consequently high uncertainties especially when the object is only marginally resolved.

Another important point is that $\sigma$ and $r_{spec}$ encode different pieces of information about the gas kinematics but are extracted from the same data, thus avoiding possible biases which might affect the dynamical mass estimate when, e.g., using the continuum emission to estimate sizes.
We conclude this section with a remark: as for the classical virial mass estimator and dynamical mass estimates from full dynamical modeling, 
with our spectroastrometric virial mass estimator we implicitly assume that the gas kinematics is that of a rotating disk.
As such, we can apply it only to objects that show gas kinematics consistent with rotation, or which are spatially unresolved.

\subsection{Calibration of the spectroastrometric estimator}\label{s52}

The next step is to calibrate the spectroastrometric mass estimator and verify its reliability with real data. We thus compare the estimator predictions with the more accurate estimates from full dynamical modeling of our $z\sim3$ galaxies from the AMAZE and LSD samples and of the $z\sim2$ subsample of high S/N rotating disks from the SINS survey \citep{Cresci:2009}.

\begin{table*}
  \caption[!ht]{}
  \label{tab1}
  \centering
  \begin{tabular}{l c c c c c c c c}
    \hline
    \hline
    \noalign{\smallskip}
     Object$^{(1)}$ & Sample & z & $FWHM\,[kms^{-1}]^{(2)}$ & $R_c\,[\arcsec]^{(3)}$ & $\Vir_{vir}\,^{(5)}$ & $r_{spec}\,[\arcsec]^{(4)}$ & $\Vir_{spec}\,^{(5)}$ & $M_{dyn}sin^2i\,^{(5)}$  \\
    \noalign{\smallskip}
    \hline
    \hline
    \noalign{\smallskip}
    SSA22a-M38 &  AMAZE &  $3.288$ & $489\pm36$ &  $0.51\pm0.05$ & $11.38\pm0.06$ & $0.29\pm0.02$ & $11.08\pm0.16$ & $11.33\pm0.02$\\
    \noalign{\smallskip}
    SSA22a-C16 &  AMAZE &  $3.065$ &   $274\pm24$ &  $0.35\pm0.02$ & $10.82\pm0.08$ & $0.09\pm0.01$ & $10.05\pm0.20$ & $10.19\pm0.02$\\
    \noalign{\smallskip}
    CDFS-2528 &  AMAZE &  $3.689$ &   $295\pm47$ &  $0.31\pm0.08$ & $10.83\pm0.13$ & $0.13\pm0.02$ & $10.24\pm0.3$ & $10.59\pm0.05$\\
    \noalign{\smallskip}
    SSA22a-D17 &  AMAZE &  $3.079$ &   $514\pm108$ &  $0.17\pm0.11$ & $11.24\pm0.28$ & $0.08\pm0.01$ & $10.57\pm0.43$  & $10.52\pm0.06$\\
    \noalign{\smallskip}
    CDFA-C9 &  AMAZE &  $3.219$ &   $320\pm25$ &  $0.19\pm0.09$ & $10.85\pm0.22$ & $0.042\pm0.003$ & $9.87\pm0.16$  & $9.81\pm0.03$ \\
    \noalign{\smallskip}
    CDFS-9313 &  AMAZE &  $3.652$ &   $264\pm25$ &  $0.22\pm0.09$ & $10.68\pm0.19$ & $0.016\pm0.009$ & $9.27\pm0.58$  & $9.14\pm0.04$ \\
    \noalign{\smallskip}
    CDFS-9340 &  AMAZE &  $3.652$ &   $150\pm46$ &  $0.19\pm0.12$ & $10.18\pm0.30$ & $0.09\pm0.03$ & $9.55\pm0.56$  & $10.1\pm0.1$ \\
    \noalign{\smallskip}
    3C324-C3 &  AMAZE &  $3.282$ &   $303\pm51$ &  $0.50\pm0.06$ & $10.99\pm0.09$ & $0.09\pm0.02$ & $10.13\pm0.42$ & $9.84\pm0.12$  \\
    \noalign{\smallskip}
    CDFS-14111 &  AMAZE &  $3.596$ &   $217\pm26$ &  $0.21\pm0.08$ & $10.47\pm0.17$ & $0.046\pm0.007$ & $9.56\pm0.24$  & $9.59\pm0.04$ \\
    \noalign{\smallskip}
    CDFS-16767 &  AMAZE &  $3.615$ &   $281\pm44$ &  $0.16\pm0.10$ & $10.63\pm0.27$ & $0.05\pm0.01$ & $9.80\pm0.34$  & $9.76\pm0.11$\\
    \noalign{\smallskip}
    Q0302-C131 &  LSD &  $3.240$ &   $176\pm22$ &  $0.07\pm0.02$ & $9.93\pm0.38$ & $0.044\pm0.007$ & $9.36\pm0.24$  & $8.98\pm0.07$ \\
    \noalign{\smallskip}
    \hline
  \end{tabular} 
\tablefoot{\tablefoottext{1}{ The spatial resolution of observations was $\sim0.6\arcsec$ (FWHM). }\tablefoottext{2}{Corrected for the instrumental spectral broadening  by subtracting in quadrature the spectral resolution. }\tablefoottext{3}{Corrected for the instrumental beam smearing by subtracting in quadrature the PSF FWHM. }\tablefoottext{4}{ No correction for instrumental beam smearing, the spectroastrometry technique enables to make distances measurements down to $\sim1/10$ of the spatial resolution (see text). }\tablefoottext{5}{Masses are in unity of $\log(M/M_{\sun})$.}}
\end{table*}

In the right panel of Fig.~\ref{fig04} we compare the spectroastrometric virial product ($\Vir_{spec}=FWHM^2r_{spec}/G$) and the dynamical mass ($M_{dyn}$) resulting from complete dynamical modeling of the galaxies analyzed in \cite{Gnerucci:2011}. The values for the dynamical mass of the best fit models and for the spectroastrometric mass estimator of the various objects analyzed in this paper are also reported in table \ref{tab1}. We also selected $8$ objects with high S/N, ``rotating disk'' gas kinematics and robust dynamical modeling from the SINS sample \citep{Cresci:2009}.
There is a tight linear correlation between $\Vir_{spec}$ and $M_{dyn}sin^2i$.

By fitting the linear relation to the data
 \begin{equation}
\log(M_{dyn}sin^2i) = \log(\Vir_{spec})+log(f_{spec}) 
\label{e5}
\end{equation}
we obtain  $\log f_{spec}=0.02\pm0.05$, i.e.~$f_{spec}=1.0^{+0.1}_{-0.1}$ with a residual dispersion of $0.15$~dex.  The right panel of Fig.~\ref{fig04} should be compared with the left panel  where we have  plotted the similar relation but using  the ``classical'' virial product $\Vir_{vir}$ instead of the  spectroastrometric virial product $\Vir_{spec}$. With our new dynamical mass estimator the relation with $M_{dyn}sin^2i$ becomes tighter (the rms of the residuals is decreased by $0.15$~dex)
and, most importantly, becomes linear compared to the non-linear relation with the classical virial estimator.
In conclusion, compared to spectroastrometric estimates, ``classical'' virial masses are more biased by systematic effects (especially in the low mass range) and are less accurate proxies of the dynamical mass.

Finally, using the fitted  calibration factor, we can obtain the value of the inclination-independent dynamical mass for an object based on the spectroastrometric mass estimator as
 \begin{equation}
M_{dyn}sin^2i=1.0^{+0.1}_{-0.1}\, \Vir_{spec}
\label{e6}
\end{equation}
with a systematic error of $\pm0.15$~dex.

We have performed a set of simulations in order to understand how the spectroastrometric mass estimator is affected by various parameters (i.e. the object's own dynamical features or the instrumental setup) and to confirm the results obtained with our data sample. These simulations, which are presented in detail in Appendix \ref{sa1}, indicate that

\begin{itemize} 
\item the spectroastrometric dynamical mass estimator is independent of the spatial resolution of the data (within the ``observational'' errors) for a variation of the PSF FWHM in the range $0.1\arcsec-1.0\arcsec$);
\item the estimator depends on the disk inclination like the dynamical mass, following a $sin^2i$ law;
\item the estimator allows one to recover the dynamical mass with a rms dispersion of $0.3$~dex when using for the simulations a S/N, disk scale length and dynamical mass values similar to those of the real data. 
\item for an exponential disk model $r_{spec}$ correspond on average to a fraction of $\sim0.75$ of the disk scale length, although this relation depends on the assumed disk profile.
\end{itemize}

\section{Application of the spectroastrometric estimator to the AMAZE and LSD complete samples}\label{s6}

In this section we apply our calibrated estimator to the galaxies from the AMAZE and LSD samples where a dynamical mass estimate from full dynamical models has not been possible.
Our estimator can be applied only to objects with rotating disk kinematics. Therefore we exclude all galaxies which are spatially resolved and show complex kinematics inconsistent with a rotating disk. Clearly, we include all the objects which are unresolved and/or possess low S/N because in those cases we have no hints about their kinematics. This selection is the same that would be applied if using the classical virial method.

As previously observed in \cite{Gnerucci:2011} we implemented a simple but quantitative method to identify objects with gas kinematics consistent with disk rotation based on classifying a galaxy as ``rotating'' or ``non-rotating'' according to whether its velocity map is well fitted by a plane. Using this method we identified a subsample of 11 ``rotating'' objects and 6 ``unclassified'' objects. Therefore we apply the spectroastrometric estimator to these ``rotating'' and ``unclassified'' objects.

We first calculate $\Vir_{spec}$ and then convert it to $M_{dyn}\,sin^2i$ as explained in the previous section. We assume an average disk inclination of $i=60^{\circ}$ (the mean inclination over a population of randomly oriented disks) to finally obtain the dynamical mass.
It was not possible to derive the dynamical mass estimator $M_{spec}$ for all objects. In same cases the S/N of the spectrum was too low, in other cases the galaxies were extended and too clumpy for a reliable estimate of the red and blue centroids. Finally, we obtain a total of $16$ dynamical mass estimates ($14$ from the AMAZE sample and $2$ from the LSD sample) ranging from from $2.5\times10^7M_{\sun}$ to $2.8\times10^{10}M_{\sun}$ with a mean value of $1.3\times10^9M_{\sun}$. 

\begin{figure}[!ht]
  \centering
  \includegraphics[width=0.99\linewidth]{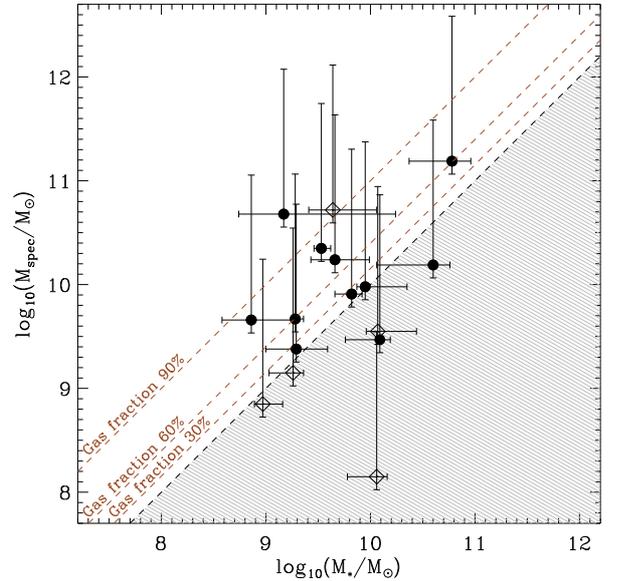}
  \caption{Spectroastrometric vs stellar mass  for AMAZE and LSD objects. Filled circles and open diamonds corresponds to objects classed as ``rotating'' and ``unclassified'' respectively, and the spectroastrometric masses have been computed for a disk inclination of $60\deg$. The  vertical bars show the effect on $M_{spec}$ of extreme disk inclinations of $90^{\circ}$ and $10^{\circ}$ (lower and upper values respectively). The gray area is the region of unphysical $M_{\star}>M_{spec}$. Brown dashed lines represents respectively the $30\%$, $60\%$ and $90\%$ gas fraction loci. Only the ``unclassified'' objects have dynamical masses  ($M_{spec}$) significantly lower than the corresponding stellar masses.}
  \label{fig07}
  \end{figure}

The values obtained are presented in figure \ref{fig07} where we compare our spectroastrometric mass estimates with the stellar masses of the ``rotating'' or ``unclassified'' objects in the AMAZE and LSD samples (we use the spectroastrometric estimates even for the galaxies where a full dynamical estimate was possible). The stellar masses for this subsample of the AMAZE and LSD samples are estimated from standard broad-band SED fitting (see Sect. 6 of \citealt{Maiolino:2008}; and Sect. 4.3 of \citealt{Mannucci:2009}), and scaled to a \cite{Chabrier:2003} IMF dividing the $M_\star$ values calculated using a Salpeter IMF for a factor $\sim1.7$ \citep{Pozzetti:2007}.
We note that we estimate $M_{dyn}sin^2i$ and assume $i=60\deg$. Therefore, in comparing spectroastrometric and stellar masses we have to take into account the effect of the disk inclination on the dynamical masses. In figure \ref{fig07} we plot vertical bars for the spectroastrometric dynamical mass corresponding to inclinations of $90^{\circ}$, $60^{\circ}$, $10^{\circ}$ (lower, central and upper value respectively).

The comparison between dynamical masses for $i=60^{\circ}$ and stellar masses, indicates that about one fourth of the dynamical masses are unphysically underestimated with respect to stellar masses. 
This fraction is mainly composed of ``unclassified'' objects. Indeed, 2 out of 5 ``unclassified'' objects have $M_{dyn}$ lower than $M_{*}$ for more than $0.2$dex while this is true for only 2 out of the 11 ``rotating'' objects.

Apart for the non-negligible systematic errors affecting stellar masses, the causes behind an underestimate of dynamical masses should be ascribed either to the unknown  disk inclination and/or to the possible contribution of  non-rotational motions in ``unclassified'' objects. 

Regarding the first effect, we assume a mean inclination of $60^\circ$ in our estimates, but we cannot verify this assumption. The vertical bars plotted in the figure identify the $10-90\deg$ inclination range and show that, when taking into account inclination effects, only one object is not consistent with the condition $M_{\star}<M_{dyn}$.
Regarding the second effect, we note that the largest discrepancies are for the ``unclassified'' objects for which the assumption of rotating gas disks cannot be verified. Indeed the presence of non-rotational motions could significantly affect the spectroastrometric mass estimate.
We will discuss in more detail all the effects that can bias our dynamical mass estimate in the next section.

\section{Discussion and conclusions}\label{s7}

We have presented a new method to estimate galaxy dynamical masses based on the spectroastrometry technique which is an improvement over ``classical'' virial mass estimates. This method is potentially very useful for its application to high z galaxies where the intrinsic spatial resolution and signal to noise ratio of the observations is often too poor to allow for the more reliable analysis based on spatially resolved kinematical modeling.

The principal advantage of spectroastrometry is its capability of overcoming the limit of the spatial resolution of the observations and this feature is important in our application of the technique. We estimated the characteristic dimensions of line emission regions down to $0.04\arcsec$ with a $>3\sigma$ significance; since the average spatial resolution is $\sim0.6\arcsec$,  the spectroastrometry has allowed us to directly probe spatial scales down to $\sim1/15$ of the spatial resolution. Our measurements do not rely on the deconvolution techniques which are used in the more conventional analysis.

We have shown that our spectroastrometric dynamical mass estimates correlate well with dynamical mass estimates from full dynamical modeling, better than classical virial mass estimates (Figs. \ref{fig01} and \ref{fig02}). We can conclude that the characteristic radius inferred with spectroastrometry has not only a better accuracy, but it is also more reliable compared to the ``classical'' estimate. An additional advantage is that a continuum image of the galaxy is not needed, but it is possible to  measure velocities and dimensions only from gas emission lines.

There is a fundamental requirement/assumption for spectroastrometric dynamical mass estimates as well as for classical virial ones: the gas kinematics must be that of a rotating disk. Only then is the shift between the red and blue image light centroids connected with disk rotation. 
In fact, if the gas kinematics is that of a rotating disk, the blue- and the red-shifted gas emission come from spatially distinct regions (i.e.~the two halves of the disk) and the corresponding light centroids will be shifted. In contrast, for a non-rotating disk, the blue- and red-shifted gas emission could be spatially mixed, the shift between the respective centroids will be smaller and the connection with the galaxy dynamical mass via the rotating disk model would no longer be possible.
Of course, the assumption of a rotating disk needs to be valid also to have a connection between the observed line width and the galaxy dynamical mass.  In this case, the line width is simply due to unresolved rotation. Conversely, any important non-rotational contribution to the line width will spuriously increase our mass estimate. This effect is present and unavoidable also in ``classical'' virial mass estimates.

To obtain a reliable dynamical mass estimate, it is important to verify that the object exhibits rotating disk gas kinematics. For the purposes of this paper we adopted the simple method presented in \cite{Gnerucci:2011} based on the $\chi^2$ statistics to identify smooth velocity gradients in the line-of-sight velocity map of the object (a smooth velocity gradient is in fact, for this low S/N and poor spatial resolution object, a good signature of a rotating disk kinematics). We remark that this simple method is used principally to discard objects inconsistent with disk rotation rather than to analyze the consistent ones. In fact our spectroastrometric mass estimator is mainly useful for low S/N and poorly resolved objects and in such cases even a good measure of the $\chi^2$ statistic to assess the presence of a smooth velocity gradient is difficult to obtain.

The easiest application of the method is for objects which do not show a complex morphology (e.g.\ with many knots of emission). This is often the case for high-z galaxies which are mostly barely resolved.
When the object is well resolved and the kinematics is that of a rotating disk, then a full modeling is possible. Therefore the spectroastrometric mass estimator is ideal to apply to sources which, for lack of S/N or spatial resolution, cannot be analyzed with standard kinematical models.

To calibrate the spectroastrometric mass estimator and verify its reliability we applied it to 19 high redshift galaxies, whose SINFONI spectra had been already analyzed and modeled: 11 $z\sim3$ objects from the AMAZE and LSD samples and 8 $z\sim2$ objects from the SINS sample. For each galaxy we compared the estimator prediction with the dynamical mass value resulting from the spatially resolved kinematical modeling. We observed a tight correlation between our estimator prediction and model dynamical mass values (dispersion of $0.15$~dex), indicating that the spectroastrometric estimator is reliable. From this comparison we also calibrated it, obtaining 
\begin{eqnarray}
M_{spec} sin^2i & = & (1.0\pm 0.1)\, 2.3\times 10^9\, M_\odot \left(\frac{FWHM}{100\, km/s}\right)^2 \left(\frac{r_{spec}}{1\,kpc}\right) \nonumber\\
 & & (\pm 0.15\, dex - \mathrm{systematic})
\end{eqnarray}
where $V$ is the line FWHM, and $r_{spec}$ is half the distance between the red and blue light centroids.
Such calibration, and the confidence of the reliability of the estimator, can of course be strengthened by enlarging the sample of galaxies for which both the spectroastrometric and standard dynamical mass are measured, extending the range of mass and redshift of the sources.
The classical virial mass estimates are a much worse proxy of the true dynamical mass (non-linear relation and $0.3$~dex dispersion).
We remark that we used  3 different samples of objects at different redshifts for this calibration,  obtaining a tight correlation over a full two orders of magnitude in $M_{dyn}sin^2i$.

As an example of its application, we finally applied our estimator to 16 objects of the AMAZE and LSD samples showing that it is possible to estimate a wide range of masses from $1.3\times10^8 M_{\sun}$ to $1.5\times10^{11} M_{sun}$ with a mean value of $7.1\times10^9 M_{sun}$. We then compared spectroastrometric and stellar masses,  $M_{dyn}$ vs. $M_{\star}$, 
showing the derived masses are consistent with the condition $M_{\star}<M_{dyn}$, once the (unknown) disk inclination is taken into account. Indeed, the estimator presented in the above equation is actually $M_{dyn}sin^2i$ , and the inclination dependence can be corrected by assuming an average value of $i=60\deg$, if no other information is available.

Our spectroastrometric estimator is given by the product of the spectroastrometric radius $r_{spec}$ with the line FWHM squared, therefore the principal biases to our mass estimate will lie in these two quantities.

A first bias can originate from the unverified assumption of rotating disks. In this case, as previously observed, the $r_{spec}$ estimate will be smaller because of the spatial mixing of blue and red-shifted gas emission leading to a underestimated dynamical mass. In our $z\sim3$ sample, the low S/N of data can be responsible, in some cases, for the selection of intrinsically non-rotating galaxies (especially for the ``unclassified'' objects) introducing spurious low-mass points in the plot of Fig.~\ref{fig07}.

Another effect that could bias the dynamical mass estimate is the presence of low-inclination (face-on) rotating disks. In this case the line FWHM will be lower because of the projection along the line of sight. Also $r_{spec}$ could be biased toward small values because the lower surface brightness of the outer regions of a face-on disk can push the estimated position of the blue and red centroids toward the disk center. Also in this case we will obtain a lower dynamical mass estimate mainly because of the assumed inclination of $60^\circ$. We note that in a high-z sample like ours, the cases with lower S/N and spatial resolution (i.e.~``unclassified'' objects) can correspond to low inclination objects. In fact lower surface brightness and a velocity gradient consistent with zero can be indications of a face-on disk. Hence an inclination smaller than $60^\circ$  might be more representative for these objects. Unfortunately it is very difficult to assess this effect more quantitatively.  

Nonetheless it is important to observe that in the case of a low mass estimate it is not possible to distinguish between the above mentioned effects because of the previously noted kinematical ambiguity for ``unclassified'' objects  (i.e.~ we cannot distinguish between low-inclination disks or non-rotating object).

Another effect that can bias our dynamical mass estimate is the presence of an important contribution to the line width from non-rotational motions. This effect rises the mass estimate and can also produce minor biases on the $r_{spec}$ estimate. In fact, if the line profile is broadened by non rotational motions, we would insert  spurious high velocity planes from the line wings in the ``red'' and ``blue'' images of the source that can alter the centroid position in an unpredictable way.
Many authors suggests that the observed high turbulence of high z disks can provide further dynamical support and add to the dynamical mass a contribution of ``sigma-supported'' mass (see for example \citealt{Epinat:2009}). However, in our method it is not possible to identify such cases and quantify this effect because of the S/N and spatial resolution of our data and because the principal effect of a broadening of the line from non-rotational motions is to raise the dynamical mass estimate. This effect is present and undetectable also in ``classical'' virial mass estimates.

\begin{appendix}

\section{Simulations}\label{sa1}

We present here a set of numerical simulations in order to understand how the spectroastrometric mass estimator is affected by several parameters such as a galaxy's dynamical structure or the instrumental setup.

Our simulations are based on a dynamical model of a high z galaxy. The model consists of a rotating thin gas disk for which we neglect hydrodynamical effects, therefore the disk motion traces the full gravitational potential. The gravitational potential is given by an exponentially-decaying disk-like mass distribution with scale radius $r_0$ and a total mass $M_{dyn}$ at a $10kpc$ radius (this is the same definition of dynamical mass given in \cite{Cresci:2009}). The galaxy is placed at redshift $z$ and has an inclination $i$ with respect to the line of sight.
We simulate integral field spectroscopic observations of a gas emission line by setting the principal instrumental parameter (spatial and spectral resolution and x, y and wavelength binsize) to match those of SINFONI, the instrument with which the data presented in this paper has been obtained.
For a detailed description of the model refer, e.g., to appendix B of \cite{marconi:2003a}.

We simulate the presence of noise by adding to the resulting data cube normally distributed random numbers, characterized by zero mean and standard deviation $\sigma_{noise}$.  This is chosen such that spectra of the galaxy, extracted from apertures similar to those used with real data, have the same $S/N$ as observations.

Finally, we analyze these simulated data cubes in the same way as the real ones and calculate the spectroastrometric virial product ($\Vir_{spec}=FWHM^2r_{spec}/G$). The uncertainties of the ``simulated'' $\Vir_{spec}$ values depend on the uncertainties on the measured photocenters positions and line FWHM, just as in the real data.

The parameters of the baseline model are chosen to match the average characteristics of the observed objects:
\begin{itemize}
\item redshift $z=3$, 
\item characteristic radius $r_0=2kpc$, 
\item dynamical mass $M_{dyn}=10^{11}M_{\sun}$, 
\item inclination $i=60^{\circ}$, 
\item spatial resolution $0.65\arcsec$FWHM, 
\item spatial scale $0.125\arcsec \times 0.125\arcsec$,
\item spectral resolution $80km s^{-1}$FWHM, 
\item dispersion $75 km s^{-1}$. 
\end{itemize}
In the following, we will vary some of these parameters and observe how this affects the estimated dynamical mass $M_{spec}$ with respect to the ``true'' model dynamical mass $M_{dyn}$. The ranges of variation of these parameters are chosen to approximately match the distribution of the various parameters values observed in our galaxy samples.

\begin{figure}[!ht]
  \centering
  \includegraphics[width=0.99\linewidth]{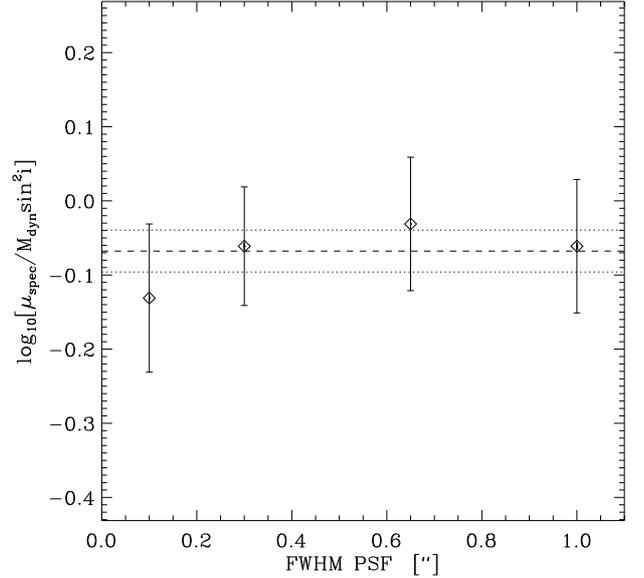}
  \caption{Simulations. Ratio between the spectroastrometric virial product $\Vir_{spec}=FWHM^2r_{spec}/G$ and the model $M_{dyn}sin^2i$ for simulations with various PSF FWHM values. The dashed line represents the mean value of the ratio. Dotted lines represent the rms dispersion of the points around the mean value.}
  \label{fig05x1}
  \end{figure}

In the first set of simulations we test the effect of varying the spatial resolution of the observations. In Fig.~\ref{fig05x1} we show the ratio (in log scale) between the model dynamical mass $M_{dyn}sin^2i$ and  the spectroastrometric virial product ($\Vir_{spec}=FWHM^2r_{spec}/G$) for various spatial resolution values. The values are perfectly consistent, confirming what we already found in \cite{Gnerucci:2010} that the spectroastrometric technique is independent of the data spatial resolution. For the baseline model only and the simulation presented in the figure, the ratio $\Vir_{spec}/M_{dyn}sin^2i$ has a mean value of $\sim-0.07$dex with a dispersion of the points of $\sim0.03$dex. We remark that the calibration factor determined in the paper represents the average $\Vir_{spec}/M_{dyn}sin^2i$ for the analyzed galaxies.

\begin{figure}[!ht]
  \centering
  \includegraphics[width=0.99\linewidth]{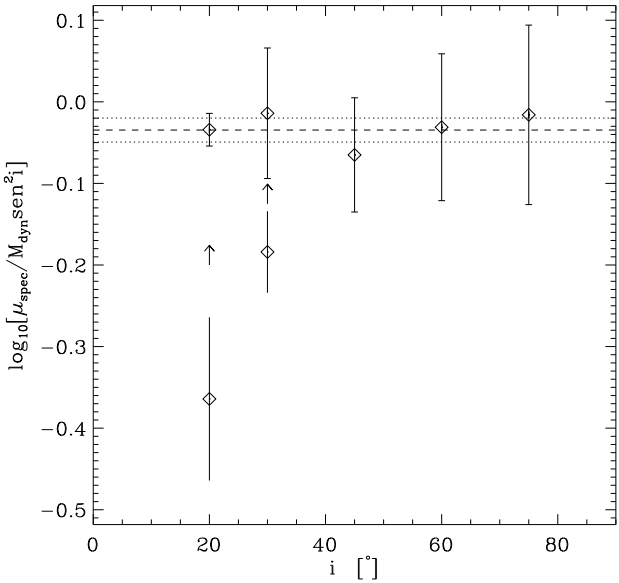}
  \caption{Simulations. Ratio $\Vir_{spec}/(M_{dyn}sin^2i)$ for simulations with various disk inclination values. At $i<45^{\circ}$ the line profiles become unresolved, thus violating the conditions under which spectroastrometry can be applied. To remedy this, the upward arrows at $i=20^{\circ}$ and $30^{\circ}$ indicate measurements for which the spectral resolution and dispersion were increased. The dashed line represents the mean value of the ratio. Dotted lines represent the rms dispersion of the points around the mean value.}
  \label{fig05x2}
  \end{figure}

In the second set of simulations we test the effect of varying the disk inclination, but keeping  all the other parameters of the basic model fixed. We expect that the spectroastrometric virial product will vary following a $sin^2i$ law due to the fact we are measuring the velocity of the gas along the line of sight; therefore, the ratio $\Vir_{spec}/M_{dyn}sin^2i$ should be independent of inclination. In Fig.~\ref{fig05x2} we show the dependence of this ratio with the inclination angle. $\Vir_{spec}/M_{dyn}sin^2i$  is systematically underestimated by $\sim0.2-0.4$dex in the simulations with $i<45^{\circ}$. This discrepancy is due  to the fact that the line FWHM becomes smaller than the instrumental spectral resolution at low inclinations and the line profile is  spectrally unresolved, thus affecting the spectroastrometric measurements. As discussed in \cite{Gnerucci:2010}, the line profiles should always be well spectrally resolved to apply spectroastrometry.
Therefore, we repeated the simulations at low-inclination increasing by a factor 2 both the spectral resolution and dispersion with respect to the baseline model: the new $\Vir_{spec}/M_{dyn}sin^2i$ values are marked by the upward arrows in the figure.
Once all the simulations have been computed with the proper spectral resolution and dispersion, the  $\Vir_{spec}/M_{dyn}sin^2i$  ratio remains fairly constant as expected. We finally remark that the problem of poor spectral resolution is not present in our data, since the line FWHM are always much larger than the instrumental spectral resolution.

  \begin{figure}[!ht]
  \centering
  \includegraphics[width=0.99\linewidth]{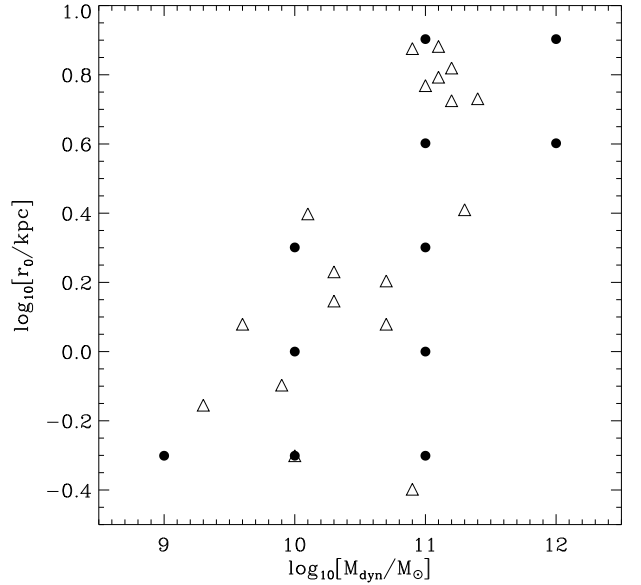}
  \caption{Distribution of dynamical masses and disk scale radii in the data sample and in the simulations. Open triangles: values observed in the data sample. Filled circles: values selected for the simulations.}
  \label{fig05x3b}
  \end{figure}

  \begin{figure*}[!ht]
  \centering
  \includegraphics[width=0.48\linewidth]{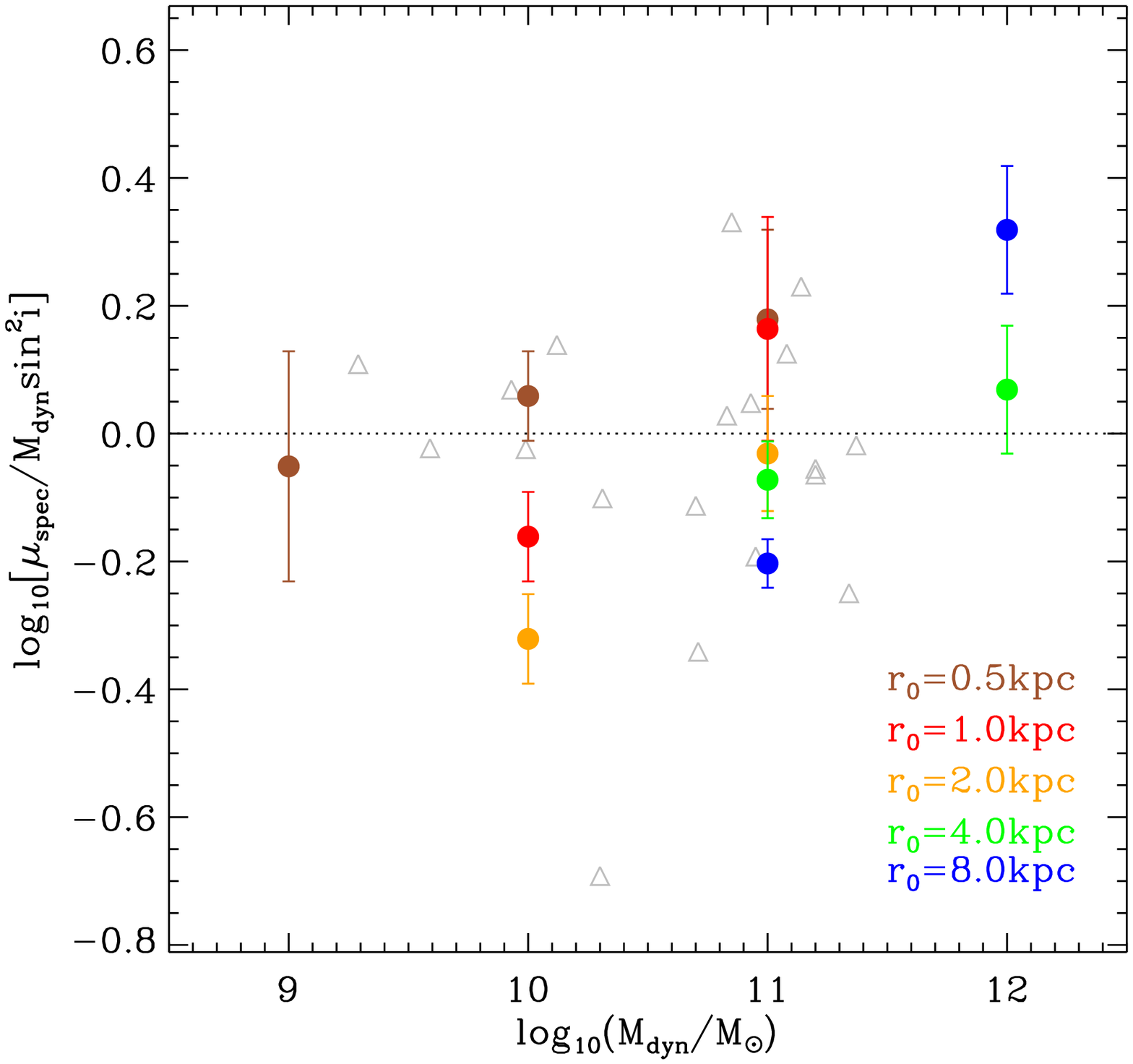}
  \includegraphics[width=0.48\linewidth]{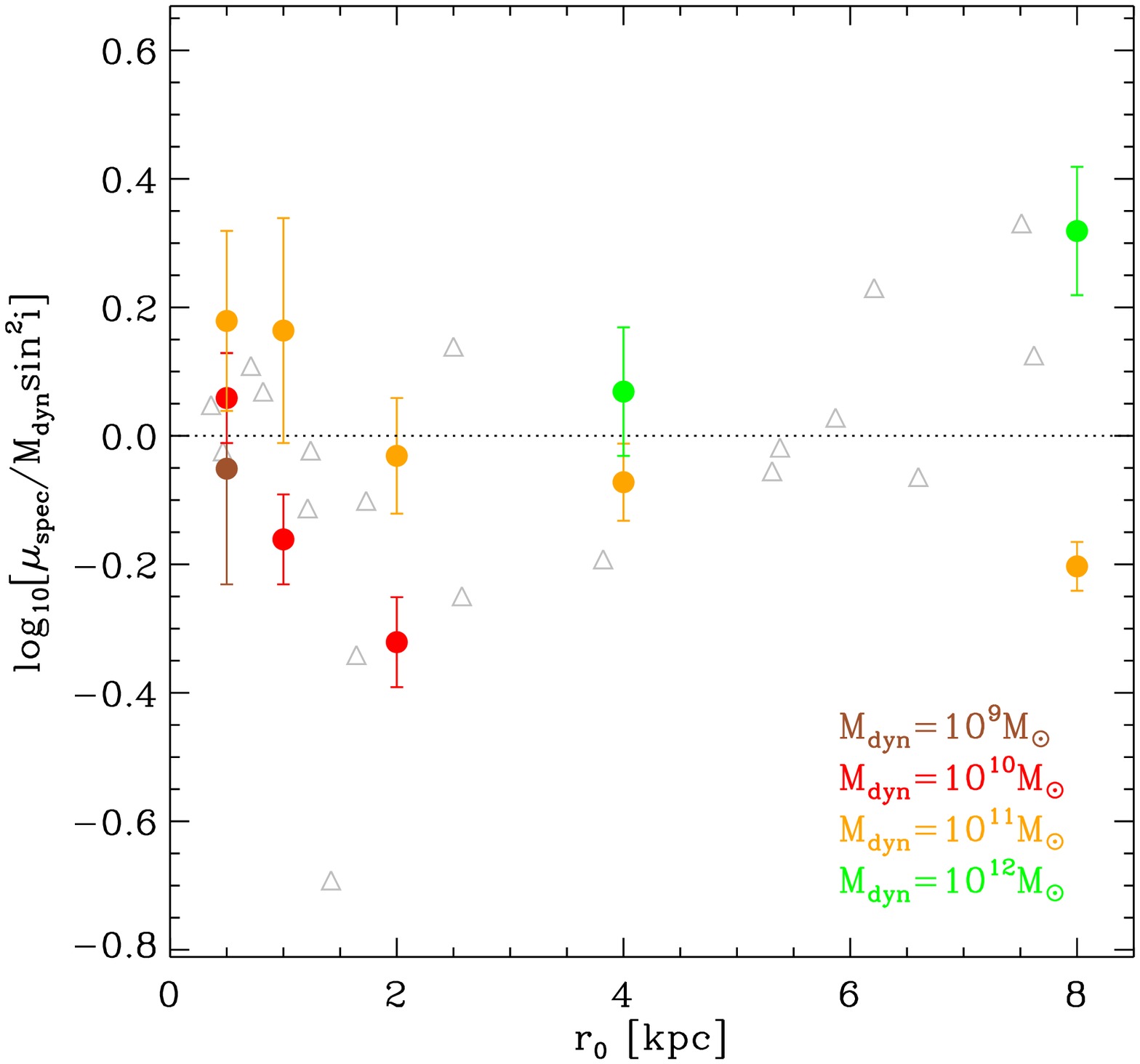}
  \caption{Simulations. Left panel: measured $\Vir_{spec}/M_{dyn}sin^2i$ values from simulations obtained with various combinations of $M_{dyn}$ and $r_0$, plotted as a function of $M_{dyn}$. Colors denote different   $r_0$ values. Right panel: same as left panel but $\Vir_{spec}/M_{dyn}sin^2i$ is now plotted as a function of $r_0$ and colors denote different $M_{dyn}$ values. In both panels gray triangles represent the measurements from real data analyzed in this paper.}
  \label{fig05x3}
  \end{figure*}

In the third set of simulations we vary the disk scale radius $r_0$ and the dynamical mass $M_{dyn}$. $r_0$ varies in the range $(0.5-8\,kpc)$ and $M_{dyn}$ in the range $(10^9-10^{12}M_{\sun})$.
The values for the simulations have been selected following a comparison with the observed $(r_0, M_{dyn})$ distribution for our sample, which is shown in Fig.~\ref{fig05x3b} (we plot the observed $(r_0, M_{dyn})$ values as open triangles and the values selected for simulations as filled circles). We note that  $r_0$ is the ionized gas scale length and $M_{dyn}$ is the dynamical mass from the full dynamical modeling of the objects analyzed by \cite{Gnerucci:2011} and used in this paper.%

In Fig.~\ref{fig05x3} we show the ratio $\Vir_{spec}/M_{dyn}sin^2i$ as a function of $M_{dyn}$  (left panel) and $r_0$  (right panel). In each panel we vary the color of the symbols to identify either $r_0$ (left panel) or $M_{dyn}$ (right panel). We also plot for comparison the measurements from the  real data (gray triangles).
Figure clearly shows that our simulations well reproduce the real measurements with a mean ratio $\Vir_{spec}/M_{dyn}sin^2i$ of $\sim-0.05$dex and a dispersion of $\sim0.3$dex around this value, consistently with the $f_{spec}$ value obtained in Sec.~\ref{s52}.

\begin{figure}[!ht]
  \centering
  \includegraphics[width=0.99\linewidth]{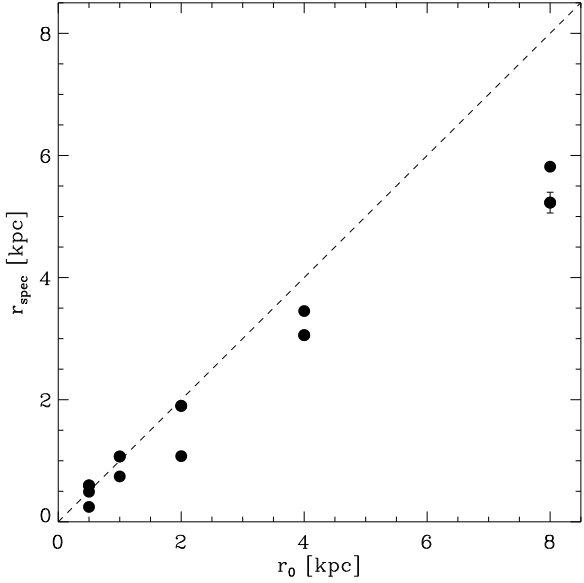}
  \caption{ Simulations. Measured $r_{spec}$ values for all the models of Figs.~\ref{fig05x3b} and \ref{fig05x3} as a function of model disk scale radius $r_0$.}
  \label{fig05x4}
  \end{figure}

Finally we compare the measured spectroastrometric characteristic radius $r_{spec}$ with the model disk scale radius $r_0$. In Fig.~\ref{fig05x4} we show this comparison for all the models plotted in Figs.~\ref{fig05x3b} and \ref{fig05x3}. From figure we can observe that $r_{spec}$ is similar to the disk scale radius $r_0$ for the smaller disks ($r_0<2kpc$) whereas for larger disks it is smaller than $r_0$ by $\sim20\%-30\%$. From the definition of $\Vir_{spec}$ (eq.~\ref{e3}) we expect that this apparent underestimate of the disk scale radius can affect in the same sense the $\Vir_{spec}$ proxy of the dynamical mass. Indeed from Fig.~\ref{fig05x3} we can not observe any evident negative bias for larger disk scale radii. This is a clear indication that the size needed in combination with the line width to obtain the mass is not accurately given by $r_0$, but by our spectroastrometric radius.

Finally, based on the simulations presented we can conclude that
\begin{itemize} 
\item the spectroastrometric dynamical mass estimator is independent of the spatial resolution of the data (within the ``observational'' errors) for a variation of the PSF FWHM in the range $0.1\arcsec-1.0\arcsec$);
\item the estimator depends on the disk inclination like the dynamical mass, following a $sin^2i$ law;
\item the estimator allows one to recover the dynamical mass with a rms dispersion of $0.3$~dex when using for  the simulations a S/N  and $(r_0, M_{dyn})$ values  similar to those of the real data. 
\item for an exponential disk model $r_{spec}$ correspond on average to $\sim0.75r_0$, although this relation depends on the assumed disk profile.
\end{itemize}

\end{appendix}
  
\begin{acknowledgements}
We would like to thank the anonymous referee for his/her constructive report on this paper. We also acknowledge financial support from the Italian National Institute for Astrophysics by INAF CRAM 1.06.09.10
\end{acknowledgements}

\bibliographystyle{aa} 

\begin{thebibliography}{50}
\expandafter\ifx\csname natexlab\endcsname\relax\def\natexlab#1{#1}\fi

\bibitem[{{Aime} {et~al.}(1988){Aime}, {Borgnino}, {Lund}, \&
  {Ricort}}]{Aime:1988}
{Aime}, C., {Borgnino}, J., {Lund}, G., \& {Ricort}, G. 1988, in European
  Southern Observatory Astrophysics Symposia, Vol.~29, European Southern
  Observatory Astrophysics Symposia, ed. F.~{Merkle}, 249--256

\bibitem[{{Baines} {et~al.}(2004){Baines}, {Oudmaijer}, {Mora}, {Eiroa},
  {Porter}, {Mer{\'{\i}}n}, {Montesinos}, {de Winter}, {Cameron}, {Davies},
  {Deeg}, {Ferlet}, {Grady}, {Harris}, {Hoare}, {Horne}, {Lumsden}, {Miranda},
  {Penny}, \& {Quirrenbach}}]{Baines:2004}
{Baines}, D., {Oudmaijer}, R.~D., {Mora}, A., {et~al.} 2004, \mnras, 353, 697

\bibitem[{{Beckers}(1982)}]{beckers:1982}
{Beckers}, J.~M. 1982, Optica Acta, 29, 361

\bibitem[{{Binney} \& {Tremaine}(2008)}]{Binney:2008}
{Binney}, J. \& {Tremaine}, S. 2008, {Galactic Dynamics: Second Edition}, ed.
  {Binney, J.~\& Tremaine, S.} (Princeton University Press)

\bibitem[{{Blumenthal} {et~al.}(1984){Blumenthal}, {Faber}, {Primack}, \&
  {Rees}}]{Blumenthal:1984}
{Blumenthal}, G.~R., {Faber}, S.~M., {Primack}, J.~R., \& {Rees}, M.~J. 1984,
  \nat, 311, 517

\bibitem[{{Bouch{\'e}} {et~al.}(2007){Bouch{\'e}}, {Cresci}, {Davies},
  {Eisenhauer}, {F{\"o}rster Schreiber}, {Genzel}, {Gillessen}, {Lehnert},
  {Lutz}, {Nesvadba}, {Shapiro}, {Sternberg}, {Tacconi}, {Verma}, {Cimatti},
  {Daddi}, {Renzini}, {Erb}, {Shapley}, \& {Steidel}}]{Bouche:2007}
{Bouch{\'e}}, N., {Cresci}, G., {Davies}, R., {et~al.} 2007, \apj, 671, 303

\bibitem[{{Chabrier}(2003)}]{Chabrier:2003}
{Chabrier}, G. 2003, \pasp, 115, 763

\bibitem[{{Christy} {et~al.}(1983){Christy}, {Wellnitz}, \&
  {Currie}}]{christy:1983}
{Christy}, J.~W., {Wellnitz}, D.~D., \& {Currie}, D.~G. 1983, Lowell
  Observatory Bulletin, 9, 28

\bibitem[{{Conselice} {et~al.}(2007){Conselice}, {Bundy}, {Trujillo}, {Coil},
  {Eisenhardt}, {Ellis}, {Georgakakis}, {Huang}, {Lotz}, {Nandra}, {Newman},
  {Papovich}, {Weiner}, \& {Willmer}}]{Conselice:2007}
{Conselice}, C.~J., {Bundy}, K., {Trujillo}, I., {et~al.} 2007, \mnras, 381,
  962

\bibitem[{{Cresci} {et~al.}(2009){Cresci}, {Hicks}, {Genzel}, {Schreiber},
  {Davies}, {Bouch{\'e}}, {Buschkamp}, {Genel}, {Shapiro}, {Tacconi},
  {Sommer-Larsen}, {Burkert}, {Eisenhauer}, {Gerhard}, {Lutz}, {Naab},
  {Sternberg}, {Cimatti}, {Daddi}, {Erb}, {Kurk}, {Lilly}, {Renzini},
  {Shapley}, {Steidel}, \& {Caputi}}]{Cresci:2009}
{Cresci}, G., {Hicks}, E.~K.~S., {Genzel}, R., {et~al.} 2009, \apj, 697, 115

\bibitem[{{Cresci} {et~al.}(2010){Cresci}, {Mannucci}, {Maiolino}, {Marconi},
  {Gnerucci}, \& {Magrini}}]{Cresci:2010}
{Cresci}, G., {Mannucci}, F., {Maiolino}, R., {et~al.} 2010, \nat, 467, 811

\bibitem[{{Davis} {et~al.}(1985){Davis}, {Efstathiou}, {Frenk}, \&
  {White}}]{Davis:1985}
{Davis}, M., {Efstathiou}, G., {Frenk}, C.~S., \& {White}, S.~D.~M. 1985, \apj,
  292, 371

\bibitem[{{Eisenhauer} {et~al.}(2003){Eisenhauer}, {Abuter}, {Bickert},
  {Biancat-Marchet}, {Bonnet}, {Brynnel}, {Conzelmann}, {Delabre}, {Donaldson},
  {Farinato}, {Fedrigo}, {Genzel}, {Hubin}, {Iserlohe}, {Kasper},
  {Kissler-Patig}, {Monnet}, {Roehrle}, {Schreiber}, {Stroebele}, {Tecza},
  {Thatte}, \& {Weisz}}]{Eisenhauer:2003}
{Eisenhauer}, F., {Abuter}, R., {Bickert}, K., {et~al.} 2003, in Society of
  Photo-Optical Instrumentation Engineers (SPIE) Conference Series, Vol. 4841,
  Society of Photo-Optical Instrumentation Engineers (SPIE) Conference Series,
  ed. {M.~Iye \& A.~F.~M.~Moorwood}, 1548--1561

\bibitem[{{Epinat} {et~al.}(2009){Epinat}, {Contini}, {Le F{\`e}vre},
  {Vergani}, {Garilli}, {Amram}, {Queyrel}, {Tasca}, \& {Tresse}}]{Epinat:2009}
{Epinat}, B., {Contini}, T., {Le F{\`e}vre}, O., {et~al.} 2009, \aap, 504, 789

\bibitem[{{Erb} {et~al.}(2006){Erb}, {Steidel}, {Shapley}, {Pettini}, {Reddy},
  \& {Adelberger}}]{Erb:2006}
{Erb}, D.~K., {Steidel}, C.~C., {Shapley}, A.~E., {et~al.} 2006, \apj, 646, 107

\bibitem[{{F{\"o}rster Schreiber} {et~al.}(2009){F{\"o}rster Schreiber},
  {Genzel}, {Bouch{\'e}}, {Cresci}, {Davies}, {Buschkamp}, {Shapiro},
  {Tacconi}, {Hicks}, {Genel}, {Shapley}, {Erb}, {Steidel}, {Lutz},
  {Eisenhauer}, {Gillessen}, {Sternberg}, {Renzini}, {Cimatti}, {Daddi},
  {Kurk}, {Lilly}, {Kong}, {Lehnert}, {Nesvadba}, {Verma}, {McCracken},
  {Arimoto}, {Mignoli}, \& {Onodera}}]{Forster-Schreiber:2009}
{F{\"o}rster Schreiber}, N.~M., {Genzel}, R., {Bouch{\'e}}, N., {et~al.} 2009,
  \apj, 706, 1364

\bibitem[{{F{\"o}rster Schreiber} {et~al.}(2006{\natexlab{a}}){F{\"o}rster
  Schreiber}, {Genzel}, {Eisenhauer}, {Lehnert}, {Tacconi}, {Nesvadba},
  {Bouch{\'e}}, {Davies}, {Lutz}, {Verma}, {Cimatti}, {Erb}, {Shapley},
  {Steidel}, {Daddi}, {Renzini}, {Kong}, {Arimoto}, {Mignoli}, {Abuter},
  {Gillessen}, {Sternberg}, \& {Gilbert}}]{Forster-Schreiber:2006a}
{F{\"o}rster Schreiber}, N.~M., {Genzel}, R., {Eisenhauer}, F., {et~al.}
  2006{\natexlab{a}}, The Messenger, 125, 11

\bibitem[{{F{\"o}rster Schreiber} {et~al.}(2006{\natexlab{b}}){F{\"o}rster
  Schreiber}, {Genzel}, {Lehnert}, {Bouch{\'e}}, {Verma}, {Erb}, {Shapley},
  {Steidel}, {Davies}, {Lutz}, {Nesvadba}, {Tacconi}, {Eisenhauer}, {Abuter},
  {Gilbert}, {Gillessen}, \& {Sternberg}}]{Forster-Schreiber:2006}
{F{\"o}rster Schreiber}, N.~M., {Genzel}, R., {Lehnert}, M.~D., {et~al.}
  2006{\natexlab{b}}, \apj, 645, 1062

\bibitem[{{F{\"o}rster Schreiber} {et~al.}(2011){F{\"o}rster Schreiber},
  {Shapley}, {Erb}, {Genzel}, {Steidel}, {Bouch{\'e}}, {Cresci}, \&
  {Davies}}]{Forster-Schreiber:2011}
{F{\"o}rster Schreiber}, N.~M., {Shapley}, A.~E., {Erb}, D.~K., {et~al.} 2011,
  \apj, 731, 65

\bibitem[{{Genzel} {et~al.}(2008){Genzel}, {Burkert}, {Bouch{\'e}}, {Cresci},
  {F{\"o}rster Schreiber}, {Shapley}, {Shapiro}, {Tacconi}, {Buschkamp},
  {Cimatti}, {Daddi}, {Davies}, {Eisenhauer}, {Erb}, {Genel}, {Gerhard},
  {Hicks}, {Lutz}, {Naab}, {Ott}, {Rabien}, {Renzini}, {Steidel}, {Sternberg},
  \& {Lilly}}]{Genzel:2008}
{Genzel}, R., {Burkert}, A., {Bouch{\'e}}, N., {et~al.} 2008, \apj, 687, 59

\bibitem[{{Genzel} {et~al.}(2006){Genzel}, {Tacconi}, {Eisenhauer},
  {F{\"o}rster Schreiber}, {Cimatti}, {Daddi}, {Bouch{\'e}}, {Davies},
  {Lehnert}, {Lutz}, {Nesvadba}, {Verma}, {Abuter}, {Shapiro}, {Sternberg},
  {Renzini}, {Kong}, {Arimoto}, \& {Mignoli}}]{Genzel:2006}
{Genzel}, R., {Tacconi}, L.~J., {Eisenhauer}, F., {et~al.} 2006, \nat, 442, 786

\bibitem[{{Gnerucci} {et~al.}(2010){Gnerucci}, {Marconi}, {Capetti}, {Axon}, \&
  {Robinson}}]{Gnerucci:2010}
{Gnerucci}, A., {Marconi}, A., {Capetti}, A., {Axon}, D.~J., \& {Robinson}, A.
  2010, \aap, 511, A19+

\bibitem[{{Gnerucci} {et~al.}(2011){Gnerucci}, {Marconi}, {Cresci}, {Maiolino},
  {Mannucci}, {Calura}, {Cimatti}, {Cocchia}, {Grazian}, {Matteucci}, {Nagao},
  {Pozzetti}, \& {Troncoso}}]{Gnerucci:2011}
{Gnerucci}, A., {Marconi}, A., {Cresci}, G., {et~al.} 2011, \aap, 528, A88+

\bibitem[{{Hopkins} \& {Beacom}(2006)}]{Hopkins:2006}
{Hopkins}, A.~M. \& {Beacom}, J.~F. 2006, \apj, 651, 142

\bibitem[{{Jones} {et~al.}(2010){Jones}, {Swinbank}, {Ellis}, {Richard}, \&
  {Stark}}]{Jones:2010}
{Jones}, T.~A., {Swinbank}, A.~M., {Ellis}, R.~S., {Richard}, J., \& {Stark},
  D.~P. 2010, \mnras, 404, 1247

\bibitem[{{Kriek} {et~al.}(2009){Kriek}, {van Dokkum}, {Franx}, {Illingworth},
  \& {Magee}}]{Kriek:2009}
{Kriek}, M., {van Dokkum}, P.~G., {Franx}, M., {Illingworth}, G.~D., \&
  {Magee}, D.~K. 2009, \apjl, 705, L71

\bibitem[{{Law} {et~al.}(2009){Law}, {Steidel}, {Erb}, {Larkin}, {Pettini},
  {Shapley}, \& {Wright}}]{Law:2009}
{Law}, D.~R., {Steidel}, C.~C., {Erb}, D.~K., {et~al.} 2009, \apj, 697, 2057

\bibitem[{{Lemoine-Busserolle} {et~al.}(2009){Lemoine-Busserolle}, {Bunker},
  {Lamareille}, \& {Kissler-Patig}}]{Lemoine-Busserolle:2009}
{Lemoine-Busserolle}, M., {Bunker}, A., {Lamareille}, F., \& {Kissler-Patig},
  M. 2009, ArXiv e-prints

\bibitem[{{Maiolino} {et~al.}(2008{\natexlab{a}}){Maiolino}, {Nagao},
  {Grazian}, {Cocchia}, {Marconi}, {Mannucci}, {Cimatti}, {Pipino}, {Ballero},
  {Calura}, {Chiappini}, {Fontana}, {Granato}, {Matteucci}, {Pastorini},
  {Pentericci}, {Risaliti}, {Salvati}, \& {Silva}}]{Maiolino:2008}
{Maiolino}, R., {Nagao}, T., {Grazian}, A., {et~al.} 2008{\natexlab{a}}, \aap,
  488, 463

\bibitem[{{Maiolino} {et~al.}(2008{\natexlab{b}}){Maiolino}, {Nagao},
  {Grazian}, {Cocchia}, {Marconi}, {Mannucci}, {Cimatti}, {Pipino}, {Fontana},
  {Granato}, {Matteucci}, {Pentericci}, {Risaliti}, {Salvati}, \&
  {Silva}}]{Maiolino:2008a}
{Maiolino}, R., {Nagao}, T., {Grazian}, A., {et~al.} 2008{\natexlab{b}}, in
  Astronomical Society of the Pacific Conference Series, Vol. 396, Astronomical
  Society of the Pacific Conference Series, ed. J.~G. {Funes} \& E.~M.
  {Corsini}, 409--+

\bibitem[{{Mannucci} {et~al.}(2007){Mannucci}, {Buttery}, {Maiolino},
  {Marconi}, \& {Pozzetti}}]{Mannucci:2007}
{Mannucci}, F., {Buttery}, H., {Maiolino}, R., {Marconi}, A., \& {Pozzetti}, L.
  2007, \aap, 461, 423

\bibitem[{{Mannucci} {et~al.}(2009){Mannucci}, {Cresci}, {Maiolino}, {Marconi},
  {Pastorini}, {Pozzetti}, {Gnerucci}, {Risaliti}, {Schneider}, {Lehnert}, \&
  {Salvati}}]{Mannucci:2009}
{Mannucci}, F., {Cresci}, G., {Maiolino}, R., {et~al.} 2009, \mnras, 398, 1915

\bibitem[{{Mannucci} \& {Maiolino}(2008)}]{Mannucci:2008}
{Mannucci}, F. \& {Maiolino}, R. 2008, in IAU Symposium, Vol. 255, IAU
  Symposium, ed. L.~K. {Hunt}, S.~{Madden}, \& R.~{Schneider}, 106--110

\bibitem[{{Marconi} {et~al.}(2003){Marconi}, {Axon}, {Capetti}, {Maciejewski},
  {Atkinson}, {Batcheldor}, {Binney}, {Carollo}, {Dressel}, {Ford}, {Gerssen},
  {Hughes}, {Macchetto}, {Merrifield}, {Scarlata}, {Sparks}, {Stiavelli},
  {Tsvetanov}, \& {van der Marel}}]{marconi:2003a}
{Marconi}, A., {Axon}, D.~J., {Capetti}, A., {et~al.} 2003, \apj, 586, 868

\bibitem[{{Mo} {et~al.}(1998){Mo}, {Mao}, \& {White}}]{Mo:1998}
{Mo}, H.~J., {Mao}, S., \& {White}, S.~D.~M. 1998, \mnras, 295, 319

\bibitem[{{Nesvadba} {et~al.}(2008){Nesvadba}, {Lehnert}, {Davies}, {Verma}, \&
  {Eisenhauer}}]{Nesvadba:2008}
{Nesvadba}, N.~P.~H., {Lehnert}, M.~D., {Davies}, R.~I., {Verma}, A., \&
  {Eisenhauer}, F. 2008, \aap, 479, 67

\bibitem[{{Nesvadba} {et~al.}(2006){Nesvadba}, {Lehnert}, {Eisenhauer},
  {Gilbert}, {Tecza}, \& {Abuter}}]{Nesvadba:2006}
{Nesvadba}, N.~P.~H., {Lehnert}, M.~D., {Eisenhauer}, F., {et~al.} 2006, \apj,
  650, 693

\bibitem[{{Nesvadba} {et~al.}(2007){Nesvadba}, {Lehnert}, {Genzel},
  {Eisenhauer}, {Baker}, {Seitz}, {Davies}, {Lutz}, {Tacconi}, {Tecza},
  {Bender}, \& {Abuter}}]{Nesvadba:2007}
{Nesvadba}, N.~P.~H., {Lehnert}, M.~D., {Genzel}, R., {et~al.} 2007, \apj, 657,
  725

\bibitem[{{Porter} {et~al.}(2004){Porter}, {Oudmaijer}, \&
  {Baines}}]{Porter:2004}
{Porter}, J.~M., {Oudmaijer}, R.~D., \& {Baines}, D. 2004, \aap, 428, 327

\bibitem[{{Porter} {et~al.}(2005){Porter}, {Oudmaijer}, \&
  {Baines}}]{Porter:2005}
{Porter}, J.~M., {Oudmaijer}, R.~D., \& {Baines}, D. 2005, in ASP Conf. Ser.
  337: The Nature and Evolution of Disks Around Hot Stars, ed. R.~{Ignace} \&
  K.~G. {Gayley}, 299--+

\bibitem[{{Pozzetti} {et~al.}(2007){Pozzetti}, {Bolzonella}, {Lamareille},
  {Zamorani}, {Franzetti}, {Le F{\`e}vre}, {Iovino}, {Temporin}, {Ilbert},
  {Arnouts}, {Charlot}, {Brinchmann}, {Zucca}, {Tresse}, {Scodeggio}, {Guzzo},
  {Bottini}, {Garilli}, {Le Brun}, {Maccagni}, {Picat}, {Scaramella},
  {Vettolani}, {Zanichelli}, {Adami}, {Bardelli}, {Cappi}, {Ciliegi},
  {Contini}, {Foucaud}, {Gavignaud}, {McCracken}, {Marano}, {Marinoni},
  {Mazure}, {Meneux}, {Merighi}, {Paltani}, {Pell{\`o}}, {Pollo}, {Radovich},
  {Bondi}, {Bongiorno}, {Cucciati}, {de la Torre}, {Gregorini}, {Mellier},
  {Merluzzi}, {Vergani}, \& {Walcher}}]{Pozzetti:2007}
{Pozzetti}, L., {Bolzonella}, M., {Lamareille}, F., {et~al.} 2007, \aap, 474,
  443

\bibitem[{{Saracco} {et~al.}(2003){Saracco}, {Longhetti}, {Severgnini}, {Della
  Ceca}, {Braito}, {Bender}, {Drory}, {Feulner}, {Hopp}, {Mannucci}, \&
  {Maraston}}]{Saracco:2003}
{Saracco}, P., {Longhetti}, M., {Severgnini}, P., {et~al.} 2003, ArXiv
  Astrophysics e-prints

\bibitem[{{Springel} {et~al.}(2006){Springel}, {Frenk}, \&
  {White}}]{Springel:2006}
{Springel}, V., {Frenk}, C.~S., \& {White}, S.~D.~M. 2006, \nat, 440, 1137

\bibitem[{{Stewart} {et~al.}(2008){Stewart}, {Bullock}, {Wechsler}, {Maller},
  \& {Zentner}}]{Stewart:2008}
{Stewart}, K.~R., {Bullock}, J.~S., {Wechsler}, R.~H., {Maller}, A.~H., \&
  {Zentner}, A.~R. 2008, \apj, 683, 597

\bibitem[{{Swinbank} {et~al.}(2007){Swinbank}, {Bower}, {Smith}, {Wilman},
  {Smail}, {Ellis}, {Morris}, \& {Kneib}}]{Swinbank:2007}
{Swinbank}, A.~M., {Bower}, R.~G., {Smith}, G.~P., {et~al.} 2007, \mnras, 376,
  479

\bibitem[{{Swinbank} {et~al.}(2009){Swinbank}, {Webb}, {Richard}, {Bower},
  {Ellis}, {Illingworth}, {Jones}, {Kriek}, {Smail}, {Stark}, \& {van
  Dokkum}}]{Swinbank:2009}
{Swinbank}, A.~M., {Webb}, T.~M., {Richard}, J., {et~al.} 2009, \mnras, 400,
  1121

\bibitem[{{Tacconi} {et~al.}(2008){Tacconi}, {Genzel}, {Smail}, {Neri},
  {Chapman}, {Ivison}, {Blain}, {Cox}, {Omont}, {Bertoldi}, {Greve},
  {F{\"o}rster Schreiber}, {Genel}, {Lutz}, {Swinbank}, {Shapley}, {Erb},
  {Cimatti}, {Daddi}, \& {Baker}}]{Tacconi:2008}
{Tacconi}, L.~J., {Genzel}, R., {Smail}, I., {et~al.} 2008, \apj, 680, 246

\bibitem[{{Tacconi} {et~al.}(2006){Tacconi}, {Neri}, {Chapman}, {Genzel},
  {Smail}, {Ivison}, {Bertoldi}, {Blain}, {Cox}, {Greve}, \&
  {Omont}}]{Tacconi:2006}
{Tacconi}, L.~J., {Neri}, R., {Chapman}, S.~C., {et~al.} 2006, \apj, 640, 228

\bibitem[{{Takami} {et~al.}(2003){Takami}, {Bailey}, \&
  {Chrysostomou}}]{takami:2003}
{Takami}, M., {Bailey}, J., \& {Chrysostomou}, A. 2003, \aap, 401, 655

\bibitem[{{Whelan} {et~al.}(2005){Whelan}, {Ray}, {Bacciotti}, {Natta},
  {Testi}, \& {Randich}}]{Whelan:2005}
{Whelan}, E.~T., {Ray}, T.~P., {Bacciotti}, F., {et~al.} 2005, in Protostars
  and Planets V, 8073--+

\end{thebibliography}

\end{document}